\pdfoutput=1
\documentclass[aps,prc,10pt,superscriptaddress,nofootinbib]{revtex4-2}
\usepackage[colorlinks=true,citecolor=blue,urlcolor=blue,linkcolor=blue,pdfstartview=FitH,bookmarksopen]{hyperref}

\usepackage{amsmath,amssymb}
\usepackage{bm}
\usepackage{comment}
\usepackage{graphicx,xcolor}

\usepackage{standalone,placeins}

\usepackage{tikz}
\usepackage{tikz-feynhand}

\newcommand\+\dagger
\newcommand\p{{\bm{p}}}
\newcommand\q{{\bm{q}}}
\renewcommand\k{{\bm{k}}}
\newcommand\x{\bm{x}}
\renewcommand\r{\bm{r}}

\renewcommand\d\partial

\newcommand\vep{\varepsilon}
\newcommand\<\langle
\renewcommand\>\rangle

\renewcommand\Im{\mathop{\mathrm{Im}}}
\DeclareMathOperator{\arccosh}{arccosh}

\usepackage[normalem]{ulem}  % \sout{old text} for strikeout

\newcommand{\diff}{\mathrm{d}}
\newcommand{\rme}{\mathrm{e}}
\newcommand{\rmi}{\mathrm{i}}

\newcommand{\cout}[1]{ \if 0 {#1} \fi }

\begin{document}

\title{
Effective field theory for weakly bound two-neutron halo nuclei:\\ corrections from neutron-neutron effective range}

\author{Davi B. Costa} 
\affiliation{Kadanoff Center for Theoretical Physics, University of Chicago, Chicago, Illinois 60637, USA}
\author{Masaru Hongo}
\affiliation{Department of Physics, Niigata University, Niigata 950-2181, Japan}
\affiliation{RIKEN iTHEMS, RIKEN, Wako 351-0198, Japan}
\author{Dam Thanh Son}
\affiliation{Kadanoff Center for Theoretical Physics, University of Chicago, Chicago, Illinois 60637, USA}

\begin{abstract}
Using an effective field-theoretical approach, we investigate the properties of weakly bound two-neutron halo nuclei (also known as Borromean nuclei) that do not support a low-energy $s$-wave core-neutron resonance.
Extending the recently formulated effective field theory for weakly bound Borromean nuclei, we incorporate corrections arising from the effective range of neutron-neutron scattering and evaluate their impact on the mean-square radii and electromagnetic response.
In particular, we compute the ratio of the matter and charge radii, the shape of the $E1$ dipole strength function, and the electric polarizability.
Our results indicate that these corrections remain numerically small when the two-neutron separation energy of the Borromean nucleus is much less than 1~MeV.
\end{abstract}

\date{March 2025}

\maketitle

\tableofcontents %just provisorily

\newpage
\section{Introduction}

Understanding the properties of nuclei at the edge of stability is one of the key goals of modern nuclear physics.
Significant experimental advances in studying neutron-rich nuclei since the 1980s~\cite{Tanihata:1985psr} have given us rich experimental data that reveal exotic behaviors (e.g., the disappearance of the traditional magic numbers), which challenge the standard picture of a nucleus.
Near the neutron drip line, one often encounters the so-called halo nuclei, which consist of a relatively compact core surrounded by an extended halo of one or more neutrons.
Among the most intriguing halo nuclei are the ``Borromean" two-neutron halo nuclei, in which the core does not form a bound state with a single neutron but does with two.
Examples of weakly bound Borromean halo nuclei include $^6$He, $^{11}$Li~\cite{Zhukov:1993aw,Kubota:2020pxo}, 
$^{19}$B~\cite{PhysRevLett.124.212503}, and $^{22}$C~\cite{Tanaka:2010,Kobayashi:2011mm,Togano:2016}.
For a comprehensive experimental overview, see also Ref.~\cite{Tanihata:2013jwa}.

Borromean nuclei exhibit several remarkable properties, presenting an intriguing nuclear many-body problem involving multiple nucleons with mass numbers in the range $A =6 \sim 29$ or more.
For instance, their matter radius is significantly larger than that of standard nuclei with a similar mass (including the core nucleus), indicating their halo structure, and knockout reactions~\cite{Corsi:2023dek} reveal the presence of the so-called ``dineutron" correlation.
Moreover, while standard nuclei exhibit giant dipole resonance in energy region $\omega = 10\sim25~$MeV, two-neutron halo nuclei exhibit a soft $E1$ dipole resonance in the much lower energy region around $\omega \sim 1$ MeV, as observed in Coulomb dissociation reactions~\cite{Ieki:1992mc,Shimoura:1994me,Zinser:1997da,Aumann:1999mb,Nakamura:2006zz}.
This resonance is expected to arise from the large spatial separation between the core and the neutrons~\cite{Hansen:1987mc}.

A crucial observation about the Borromean nuclei is that their halo structure allows for an effective description as a three-body quantum system, consisting of a compact core and two loosely bound neutrons. 
In other words, although the core itself is composed of many nucleons, low-energy properties of the Borromean nuclei can often be captured by regarding the core as a point-like object.  
This is the starting point of several theoretical approaches, including quantum models~\cite{Bertsch:1991zz,Esbensen:1992qbt,Fedorov:1994zz,Nielsen:2001hbm,Hagino:2005we,Hagino:2009sj} and a field-theoretical framework known as halo effective field theory (Halo EFT)~\cite{Bertulani:2002sz,Canham:2008jd,Frederico:2012xh,Acharya:2013aea,Hagen:2013xga,Hammer:2017tjm}.
In particular, Halo EFT regards the appearance of the Borromean nuclei as the manifestation of the Efimov effect~\cite{Efimov:1970zz,Efimov:1973awb}, well known in quantum three-body problems (see, e.g., Refs.~\cite{Braaten:2004rn,Hammer:2010kp,Naidon:2016dpf,Endo:2024cbz} for reviews). 
For that, it assumes a resonant $s$-wave interaction between the core and the neutron (i.e., a large core-neutron $s$-wave scattering length).

However, from a purely theoretical perspective, the existence of a shallow three-body bound state, comprising the core and two neutrons, does not necessarily require a core-neutron resonance.
In fact, in a simple three-body quantum mechanical model of a Borromean nucleus, where the two neutron interact via a zero-range interaction tuned to unitarity inside the potential well of a heavy core, the onset of three-body bound-state formation occurs well before the emergence of a core-neutron bound state (see Appendix \ref{app:variational} for a concrete example).
This motivates the construction of a new effective field theory (EFT) framework for describing weakly bound two-neutron halo nuclei, as proposed in Ref.~\cite{Hongo:2022sdr}.
The EFT assumes the existence of two, and only two, small energy scales: the two-neutron virtual energy $\epsilon_n$, which is related to the $s$-wave neutron-neutron scattering length $a$ $\approx -19$ fm by $\epsilon_n=\hbar^2/(m_n a^2) \approx 0.12$ MeV (where $m_n$ is the neutron mass) and the two-neutron separation energy $B\equiv S_{2n}$ of the halo nucleus (or its binding energy when regarded as a three-body system).  
While no hierarchy between $\epsilon_n$ and $B$ is assumed, all other energy scales are taken to be significantly larger.
In particular, the core-neutron scattering length is assumed to be small, in stark contrast to the key assumption of Halo EFT
\footnote{
However, this assumption of a small $s$-wave core-neutron scattering length is not valid for all two-neutron halo nuclei. 
For instance, in the cases of $^{11}$Li and $^{19}$B nuclei, low-energy $s$-wave resonances exist in the $n$--$^9$Li and $n$--$^{17}$B subsystems, respectively.  
\label{footnote-EFT-limitation}
}. 

Among the known Borromean nuclei, the carbon isotope $^{22}$C is expected to be the best candidate for a two-neutron halo nucleus described by the EFT of Ref.~\cite{Hongo:2022sdr}.
Indeed, experimental evidence suggests that the $n$--$^{20}$C scattering length is small~\cite{Mosby:2013bix}, indicating that no $s$-wave core-neutron resonance exists with an energy comparable to or smaller than $\epsilon_n$ and $B$.
Based on this EFT framework, the ratio of the charge and matter radii, as well as the shape of the $E1$ dipole strength function, was computed in Ref.~\cite{Hongo:2022sdr}.
The result for the ratio of the matter radius to the charge radius was later reproduced in Ref.~\cite{Naidon:2023fio}, making use of the crucial observation that the three-body wave function is dominated by a single Faddeev component.

Building on the results of Ref.~\cite{Hongo:2022sdr}, we can investigate various corrections to the properties of halo nuclei by incorporating possible irrelevant terms in the effective Lagrangian.
The purpose of this paper is to clarify the magnitude of one specific contribution---the correction arising from the effective range $r_0 \approx 2.75$ fm in neutron-neutron scattering [corresponding to the energy scale of $\epsilon_0\equiv\hbar^2/(m_n r_0^2)\approx 5.5\,\text{MeV}$].
Investigating the effective-range correction requires the inclusion of an additional term in the effective Lagrangian. 
By computing the correction at the first-order in $r_0$, we show that the impact of the effective-range correction on physical observables is not  significant for $^{22}$C.  
In particular, the shape of the $E1$ dipole strength function remains nearly unchanged when this correction is taken into account.
The correction to the ratio of the charge and matter radii is somewhat larger but remains numerically small for the expected small binding energy of $^{22}$C.

The paper is organized as follows.
In Sec.~\ref{sec:EFT}, we introduce a low-energy EFT that captures the universal properties of weakly bound two-neutron halo nuclei, incorporating the effective-range correction.
Using this EFT, we elucidate the universal behavior of the mean-square radii (such as charge and matter radii) in Sec.~\ref{sec:Strucuture} and the electromagnetic response (the $E1$ dipole strength function and electric polarizability) in Sec.~\ref{sec:elemag}.
Section~\ref{sec:conclusion} is devoted to our concluding remarks.
In Appendix~\ref{app:variational}, we present a simple quantum mechanical model that illustrates a key feature of our EFT.
Finally, in Appendix~\ref{app:details}, we provide technical details of our calculations.

\section{Effective field theory}
\label{sec:EFT}

In this Section, we introduce the effective field theory that captures universal properties of weakly bound halo nuclei composed of two neutrons and core. 
Our theory extends the theory proposed in Ref.\ \cite{Hongo:2022sdr} by including a term corresponding to the effective-range correction in the neutron-neutron scatterings. 
After providing an effective Lagrangian in Sec.~\ref{sec:Lagrangian}, we demonstrate the renormalization condition and its properties in Sec.~\ref{sec:renormalization}.

\subsection{Effective Lagrangian}\label{sec:Lagrangian}

The effective field theory describing a weakly bound two-neutron halo nuclei needs to incorporate the low-energy dynamics around two small energy scales: the binding energy $B$, small by assumption, and the di-neutron virtual energy $\epsilon_n = 1/(m_na^2)$, small by the large neutron-neutron scattering length $a$.
The effective Lagrangian capturing these two low-energy scales is constructed in Ref.~\cite{Hongo:2022sdr}, whose generalization, equipped with a perturbative correction induced by the effective range between neutrons $r_0$, is given by 
\begin{equation}\label{eq:bare-Lag}
  \mathcal L = 
  h_0^\+ \biggl( \rmi \d_t + \frac{\nabla^2}{2m_h} + B_0 \biggr) h_0
  + \phi^\+ \biggl( \rmi \d_t + \frac{\nabla^2}{2m_\phi} \biggr) \phi
  + g_0 ( h_0^\+ \phi d + \phi^\+ d^\+ h_0 )
  + \mathcal L_n ,
\end{equation}
where the bare field $h_0$ describes the halo nuclei with the mass $m_h = (A+2) m_n$ and the bare binding energy $B_0$, and $\phi$ describes the core nuclei with the mass $m_\phi = A m_n $ with a mass number $A$ for the core.
Here $g_0$ is a bare couling constant describing the interaction 
between the core and a dimer field $d$ composed of two neutrons. 
The Lagrangian for two neutrons $\mathcal L_n $ that encodes the effective-range correction is 
\begin{equation}\label{eq:neutron-Lag}
  \mathcal L_n = \sum_\sigma \psi^\+_\sigma 
   \biggl( \rmi \d_t + \frac{\nabla^2}{2m_n} \biggr) \psi_\sigma
  - \frac1{c_0} d^\+ d + \psi^\+_\uparrow \psi^\+_\downarrow d
  + d^\+ \psi_\downarrow \psi_\uparrow
  -\frac{r_0}{8\pi} d^\+ \biggl( \rmi \d_t + \frac{\nabla^2}{4m_n} \biggr) d,
\end{equation}
where we introduced the neutron field $\psi_\sigma$ with the spin index $\sigma=\uparrow,\downarrow$.  
The last term in Eq.~(\ref{eq:neutron-Lag}) is new compared to the Lagrangian considered in Ref.~\cite{Hongo:2022sdr}, which generates the perturbative correction proportional to the effective range $r_0$. 
As we will see shortly, $1/c_0$ and $r_0$ correspond to the neutron $s$-wave scattering length and the effective range. Throughout the paper, we measure the mass of particles with respect to the neutron mass $m_n$ by setting $m_n = 1$.

The effective Lagrangian (\ref{eq:bare-Lag}) is composed of three sectors coupled with each other: the halo sector, described by the first term in Eq.~(\ref{eq:bare-Lag}), the core sector, described by the second term, and the neutron sector described by $\mathcal L_n$ in Eq.~(\ref{eq:neutron-Lag}). 
The crucial point here is that the neutron sector is described by a nonrelativistic conformal field theory~\cite{Nishida:2007pj} deformed by one relevant operator (the term proportional to $c_0^{-1}$) and one irrelevant operator (the $r_0$ term).  
In other words, we are working around the interacting fixed point of $\mathcal L_n$, where $c_0^{-1}$ is fine-tuned to the critical value, but not around the noninteracting fixed point $c_0^{-1}=\infty$.  
Thus, the relevant power-counting scheme is the one in which one takes into account the nontrivial scaling dimensions in the nonrelativistic conformal field theory.

In short, the dimension of $\psi_\sigma$, $h$, and $\phi$ are as in free field theory: $[\psi_\sigma] = [h] = [\phi] = 3/2$, but the dimension of the dimer is $[d] = 2$ (our convention is $[\nabla] = 1$ and $[\d_t]=2$).  
The dimension of the Lagrangian density is then $[\mathcal{L}]=5$, and we find $[B_0]=[c_0^{-1}]=1$, $[g_0] = 0$, and $[r_0] = -1$.  
The last relation implies that the effective range $r_0$ induces an irrelevant dimension-6 term whose contribution to low-energy observables is suppressed.

Let us briefly comment on other corrections omitted in this paper.
We focus on the $r_0$ corrections to several properties of the halo nucleus, such as the ratio of the matter and charge radii, the $E1$ dipole strength and the electric polarizability, some of which were computed at leading order in Ref.~\cite{Hongo:2022sdr}.
The only other dimension-6 operator that can compete with the effective-range term is $\phi^\dag \psi^\dag \psi \phi$, whose coefficient is proportional to the $s$-wave scattering length $a_{cn}$ between the core and neutron.
We leave the analysis of corrections from the core-neutron scattering length for future work but note that, in the case of $^{22}$C, experimental data suggest that $a_{cn}$ may be small~\cite{Mosby:2013bix}.
As mentioned in footnote \ref{footnote-EFT-limitation}, our theory is not applicable for two-neutron halo nuclei with large $a_{cn}$ (or a shallow core-neutron $s$-wave resonance), such as $^{11}$Li and $^{19}$B.

In the case of $^6$He, the $s$-wave scattering length between a neutron and an $\alpha$-particle is not large, $a_{cn} \approx 2.6~\text{fm}$~\cite{LandoltBornstein2000:sm_lbs_978-3-540-49623-6_6},but there is a resonance in the $p$-wave channel. 
Such a resonance can be incorporated into the EFT by introducing a new field, $\chi_i$, which couples to the core and neutron via the vertex (symbolically) $g_{\chi\phi\psi}\chi_i^\+ \phi\d_i \psi + (\text{h.c.})$~\cite{Bertulani:2002sz}.  
The coupling constant $g_{\chi\phi\psi}$ is irrelevant with dimension $-1/2$. 
In physical observables related to the halo, the $\chi^\+\phi\psi$ vertex always appears at least twice in any Feynman diagram, making the effect of a $p$-wave resonance comparable in order to the $nn$ effective range or the core-neutron $s$-wave scattering length.
Since this paper focuses on the $r_0$ correction, we neglect this contribution as well.

We also note that the field $\phi$, $\psi_\sigma$ and $d$ are 
not subject to field renormalization in the nonrelativistic theory. Thus, we find the exact propagators for the neutron and core as 
\begin{align}
 \rmi G_\phi (p) 
 &= \frac{\rmi}{p_0 - \frac{\p^2}{2m_\phi} + \rmi \epsilon} ,
 \\
 \rmi G_\psi (p) 
 &= \frac{\rmi}{p_0 - \frac{\p^2}{2} + \rmi \epsilon},
\end{align}
which we express as solid lines in Feynman diagrams with $\phi$ or $\psi$.  
The full propagators for the dimer and halo will be introduced shortly, whose corresponding Feynman diagrams are represented by dashed and double lines, respectively.
The Lagrangian \eqref{eq:bare-Lag} contains two interaction vertices; the first vertex, equal to $\rmi$, describes the interaction between two neutrons and the dimer, while the second vertex, equal to $\rmi g_0$, describes the interaction between the halo, core, and dimer.
The latter vertex will be replaced by the renormalized coupling, as we will discuss shortly.

\subsection{Renormalization}\label{sec:renormalization}

We now derive the exact propagators for the dimer and halo. 
First of all, we note 
that the effective Lagrangian of Eq.~\eqref{eq:bare-Lag} contains the bare field $h_0$ as well as 
three bare couplings $c_0,B_0$ and $g_0$.
These bare quantities should be replaced by the renormalized ones. 
For that purpose, we introduce the appropriate counterterms and 
rearrange the effective Lagrangian to
\begin{equation}\label{eq:total-Lag}
 \mathcal L = 
 h^\+ \left( \rmi \d_t + \frac{\nabla^2}{2m_h} + B \right) h 
 + \phi^\+ \biggl( \rmi \d_t + \frac{\nabla^2}{2m_\phi} \biggr) \phi
 + g ( h^\+ \phi d + \phi^\+ d^\+ h )
 + \mathcal L_n 
 + (\text{counterterms}),
\end{equation}
where we introduced the renormalized field $h$ and coupling $g$ by
\begin{equation}\label{eq:bare-renormalized}
 h_0 = \sqrt{Z_h} h, \quad 
 g_0 = \frac{g}{\sqrt{Z_h}},
\end{equation}
with a field renormalization factor $Z_h$.
We here rely on the fact that there is no loop correction to the three point vertex corresponding to $g$ in the present nonrelativistic theory.
Thus, the field renormalization $Z_h$ controls the coupling renormalization.

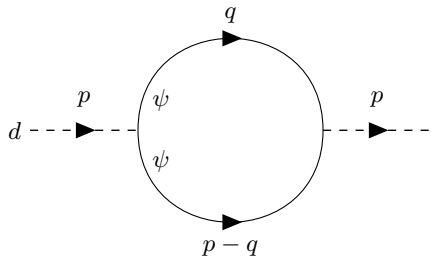
\begin{figure}[t]
  \centering
  \includegraphics[width=0.33\linewidth]{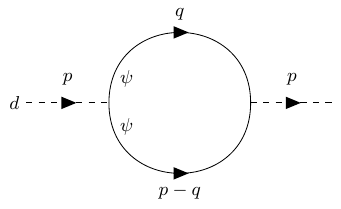}
  \cout{
  \scalebox{1}{
  \begin{tikzpicture}
   \begin{feynhand}
    \vertex (d2) at (2.6,0);
    \vertex (v2) at (1.2,0);
    \vertex (d1) at (-2.6,0);
    \vertex (v1) at (-1.2,0);
    \node at (-2.8,0) {$d$};
    \node at (-0.9,0.4) {$\psi$};
    \node at (-0.9,-0.4) {$\psi$};
    \node at (1.9,0.4) {$p$};
    \node at (-1.9,0.4) {$p$};
    \node at (0,1.5) {$q$};
    \node at (0,-1.5) {$p-q$};
    \propag [sca, with arrow=0.5] (d1) to (v1);
    \propag [with arrow=0.5] (v1) to [out=270,in=270, looseness=1.7] (v2);
    \propag [with arrow=0.5] (v1) to [out=90,in=90, looseness=1.7] (v2);
    \propag [sca, with arrow=0.5] (v2) to (d2);
   \end{feynhand}
  \end{tikzpicture}}
  }
  \caption{The self-energy for the dimer.}
\label{fig:dimer-prop}
\end{figure}

In the neutron part Eq. \eqref{eq:neutron-Lag}, we have two parameters $c_0$ and $r_0$ corresponding to 
the $s$-wave scattering length and effective range, respectively.
Computing the self-energy of the dimer field, which, in the nonrelativistic theory, is exactly given by the one-loop diagram shown in Fig.~\ref{fig:dimer-prop}, we find the full dimer propagator as 
\begin{equation}\label{dimerD}
  D^{-1}(p) = - \frac1 {4\pi} \sqrt{\frac{\p^2}4 - p_0 - \rmi\epsilon}
    +\frac1{4\pi a}
    + \frac{r_0}{8\pi} \left(  \frac{\p^2}4 - p_0 \right),
\end{equation}
where $a$ denotes the $s$-wave scattering length given by
\begin{equation}
  \frac1{4\pi a} = - \frac1{c_0} + \int\!\frac{\diff \q}{(2\pi)^3}\,
  \frac1{\q^2} \,.
\end{equation}
The bare coupling $1/c_0$ and the second term in the right-hand side 
both diverge, and they are renormalized by the finite $s$-wave scattering length $a$.
One clearly see that $a$ is indeed the $s$-wave scattering length by recalling that the dimer propagator describes the scattering problem of two incoming neutrons.
Indeed, two-neutron scattering with incoming momenta $\k$ and $-\k$ is described by the dimer propagator with $p_0=\k^2$ and $\p=0$:
\begin{equation}
  D^{-1}(p_0=\k^2,\p=0) = 
  \frac1{4\pi} \left( \frac1a + \rmi |\k| - \frac12 r_0 \k^2 \right).
\end{equation}
Comparing this result with the $T$-matrix known in quantum mechanics~\cite{Landau:QM}, we find the parameters $a$ and $r_0$ indeed match with the $s$-wave scattering length and the effective range between two neutrons.

As we have already mentioned, the term proportional to the effective range $r_0$ is a leading irrelevant term in our EFT.
We thus treat the effective-range correction perturbatively by expanding the dimer propagator with respect to $r_0$. 
In the following calculation, we express the dimer propagator as
\begin{equation}\label{eq:D-f}
  D(p) = -4\pi
 f \left(\frac{\p^2}4-p_0; a, r_0 \right)
  ,
\end{equation}
where we have defined 
\begin{equation}\label{eq:fxar0-def}
  f (x; a,r_0)
  \equiv \frac 1{\sqrt{x} - \frac1a - \frac{r_0}2 x} \,.
\end{equation}
Treating the effective range $r_0$ as a small parameter, we can expand
\begin{equation}\label{eq:def-f}
  f (x; a,r_0)
  = f(x;a) + \frac{r_0}2 x f^2(x;a) + O(r_0^2),
\end{equation}
with~\footnote{This function was denoted as $f_a(x)$ in Ref.~\cite{Hongo:2022sdr}.}
\begin{equation}\label{eq:def-fa}
  f(x;a)\equiv f(x;a,0) = \frac1{\sqrt{x} - \frac1a} \, ,
\end{equation}
which we will use in the subsequent sections.

We next discuss the renormalization of the halo part.
The vital point here is that the full halo propagator provides necessary information not only on the field and coupling renormalization [recall Eq.~\eqref{eq:bare-renormalized}], but also on binding-energy renormalization.
The exact halo self-energy $\Sigma(p)$ is given by the one-loop graph shown in Fig.~\ref{fig:halo-self-energy}. 
Performing the $q_0$-integral by closing the contour in the lower-half plane, we find the halo self-energy as
\begin{align}
 \Sigma (p) 
 &= \rmi (\rmi g)^2\!\int\! \frac{\diff^4 q}{(2\pi)^4}\, 
 \rmi D (p-q)\, \rmi G_{\phi} (q) 
 \nonumber \\
 &= - 4 \pi g^2 \!\int\! \frac{\diff \q}{(2\pi)^3}\, 
 f 
 \biggl( -p_0 + \frac{1}{4} (\p-\q)^2 + \frac{\q^2}{2m_\phi}; a, r_0 \biggr)
 \nonumber \\
 &= - 4 \pi g^2\! \int\! \frac{\diff \q}{(2\pi)^3}\, 
 f 
 \biggl( -p_0 + \frac{\p^2}{2m_h} + \frac{\q^2}{2\mu}; a, r_0 \biggr).
\end{align}
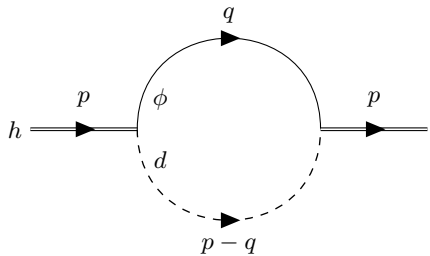
\begin{figure}[t]
 \centering
  \includegraphics[width=0.33\linewidth]{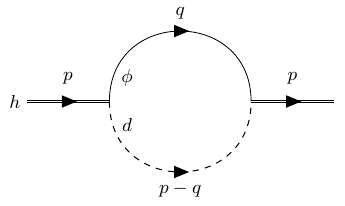}
  \cout{
  \scalebox{1}{
  \begin{tikzpicture}
   \begin{feynhand}
    \vertex (d2) at (2.6,0);
    \vertex (v2) at (1.2,0);
    \vertex (d1) at (-2.6,0);
    \vertex (v1) at (-1.2,0);
    \node at (-2.8,0) {$h$};
    \node at (-0.9,0.4) {$\phi$};
    \node at (-0.9,-0.4) {$d$};
    \node at (1.9,0.4) {$p$};
    \node at (-1.9,0.4) {$p$};
    \node at (0,1.5) {$q$};
    \node at (0,-1.5) {$p-q$};
    \propag [double, with arrow=0.5] (d1) to (v1);
    \propag [sca, with arrow=0.5] (v1) to [out=270,in=270, looseness=1.7] (v2);
    \propag [with arrow=0.5] (v1) to [out=90,in=90, looseness=1.7] (v2);
    \propag [double, with arrow=0.5] (v2) to (d2);
   \end{feynhand}
  \end{tikzpicture}}
  }
  \caption{The self-energy for the halo.}
  \label{fig:halo-self-energy}
\end{figure}
To derive the last line, we performed a shift $\q \to \q + \frac{m_\phi}{m_h}$ 
and defined the reduced mass of the core-dimer system by
\begin{equation}
 \mu \equiv \frac{2m_\phi}{m_h} = \frac{2 A }{A+2}.
\end{equation}
where $A$ is the mass number of the core.

As a result, we find the exact inverse halo propagator $G_h^{-1} (p)$ as
\begin{equation}
 G_h^{-1} (p)
 = Z_h \left( p_0 - \frac{\p^2}{2m_h} + B_0 \right)
 + 4 \pi g^2\! \int\! \frac{\diff \q}{(2\pi)^3} \,
 f \!
 \left( -p_0 + \frac{\p^2}{2m_h} + \frac{\q^2}{2\mu};a,r_0 \right).
\end{equation}
Recalling the definition of $f(x;a,r_0)$ in Eq.~\eqref{eq:def-f}, 
we find the $\q$-integral in this equation diverges in the UV: the leading term quadratically, and the $r_0$-correction term cubically~\footnote{If one does not expand $f(x;a,r_0)$ over $r_0$, the integral formally diverges only linearly, but the integrand then has an unphysical singularity at a large value of its argument.}.
Regularizing the UV divergence by, e.g., introducing a UV cutoff $\Lambda$, we perform the renormalization relying on the on-shell renormalization scheme 
\begin{align}
 G_h^{-1} (p_0,\bm{0}) \big|_{p_0 = - B} &= 0 , 
 \\
 \frac{\d}{\d p_0} 
 G_h^{-1} (p_0,\bm{0}) \big|_{p_0 = - B} &= 1 ,
\end{align}
which results in the following renormalization conditions:
\begin{align}
  Z_h (B_0 - B ) 
  + 4 \pi g^2\! \int\! \frac{\diff \q}{(2\pi)^3}\, 
  f  ( B_{\q}; a,r_0 )
  &= 0 , 
  \\
  Z_h - 4 \pi g^2 \!\int\! \frac{\diff \q}{(2\pi)^3}\, 
  f^{\prime}  ( B_{\q}; a,r_0 )
  &= 1,
  \label{eq:Zh}
\end{align}
with 
\begin{equation}\label{eq:Bq}
   B_{\q} \equiv B + \frac{\q^2}{2\mu}\,.
\end{equation}
Here and thereafter, derivatives always act on the first variable in $f(x;a,r_0)$ and $f(x;a)$, i.e., $f'(x;a,r_0)\equiv \d_x f(x; a,r_0)$, etc.
The first condition renormalizes the diverging integral and the bare binding energy with the physical binding energy $B$, 
and the second condition determines the field renormalization $Z_h$. 

Recalling Eq.~\eqref{eq:bare-renormalized}, 
we find the relation between the bare and renormalized coupling as 
\begin{equation}
 g_0^2 = \frac{g^2}{1 + 4 \pi g^2\! \int\! \frac{\diff \q}{(2\pi)^3}\, 
 f^{\prime} ( B_{\q}; a,r_0)}.
\end{equation}
Taking the derivative of this result
after expanding $f^{\prime} ( B_{\q}; a,r_0)$ 
with respect to $r_0$, we find the beta function for the coupling $g$,
\begin{equation}
 \beta (g) 
 \equiv \frac{\d g}{\d \ln E}
 = \frac{2}{\pi} 
   \left( \frac{A}{A+2} \right)^{3/2}
   g^3 
   \left[1 + \frac{r_0}{a} + O \left( \left( \frac{r_0}{a} \right)^2 \right) \right] ,
\end{equation}
where $E$ denotes the energy scale. 
Solving this renormalization group equation, we find the running coupling constant 
\begin{equation}
 g^2 (E) = \frac{\pi}{4} \left( \frac{A}{A+2} \right)^{3/2}
 \left[1 + \frac{r_0}{a} + O \left( \left( \frac{r_0}{a} \right)^2 \right) \right]
 \frac{1}{\ln \frac{E_0}{E}}, 
\end{equation}
with the energy at the Landau pole $E_0$.

The existence of the running coupling constant means that we need to fix its value at certain energy scale using, e.g., experimental data to predict absolute values of experimental observables.
In that sense, the low-energy EFT needs another input to obtain theoretical predictions in addition to the binding energy of the halo nuclei $B$, and the scattering length $a$ and effective range $r_0$ between two neutrons.
However, there is another way to extract the universal prediction by computing the ratio of theoretical calculation, which cancels the overall running coupling dependence.
In the subsequent section, we investigate the mean-square radii and electromagnetic response of the halo nuclei relying on the latter prescription.

\section{Mean-square radius}
\label{sec:Strucuture}

In this section, we investigate the effective-range correction of charge, dineutron, and matter mean-square radii of the halo nuclei.
In Sec.~\ref{sec:Geometry}, we first review a field-theoretical way to compute mean-square radii with useful relations.
After deriving the analytic formula on the charge radius in Sec.~\ref{sec:Charge-radius},
we present the neutron and matter radii in Sec.~\ref{sec:Neutron-Matter-radius}.
In Sec.~\ref{sec:discusion-rms}, we discuss magnitudes of the derived effective-range corrections.

\subsection{Form factor and mean-square radius}
\label{sec:Geometry}

As an indicator of the unique structure of halo nuclei, we focus on the root-mean-square (rms) radii of charge, dineutron, and matter distributions, denoted by $\langle r_a^2 \rangle$, where $a = c, n, m$, respectively.
To calculate these, recall that each rms radius can be obtained from the corresponding form factor $F_a (\k)$, which is the Fourier transform of the density distribution $\rho_a (\x)$ in real space. 
Specifically, by expanding the form factor for small momentum $\k$, we find:
\begin{equation}
 F_a (\k) = \int\! \diff \x\, \rme^{-\rmi \k \cdot \x} \rho_a (\x)
 = Q_a - \frac{1}{6} \k^2 \langle r_a^2 \rangle 
 + O(\k^4),
\end{equation}
where the total number $Q_a$ and the rms radius $\langle r_a^2 \rangle$ are defined as
\begin{align}
 Q_a &= \int\! \diff \x\, \rho_a (\x),
 \\
 \langle r_a^2 \rangle &= \int\! \diff \x \,\x^2 \rho_a (\x),
\end{align}
We note that the each total number is identified as $Q_c = Z$, $Q_n =2$, and $Q_m = A+2$, where $Z$ and $A$ represent the atomic number and mass number of the core in the halo nucleus, respectively.
The rms radius can thus be determined from the coefficient of the $\k^2$ term in the expansion of the form factor.

In the field-theoretical formulation, the form factor $F_a (\k)$ is diagrammatically represented by a three-point vertex involving an incoming and outgoing on-shell halo and a virtual ``photon" carrying spatial momentum $\k$ as
\vspace{30pt}
\begin{equation}
 F_a (\k)
 =
 \parbox{4.0cm}{\vspace{-34pt}\includegraphics[width=4.0cm]{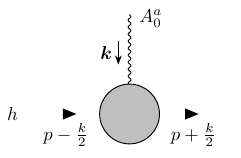}}
 \cout{
 \scalebox{0.9}{
  \begin{tikzpicture}[baseline=(o.base)]
   \begin{feynhand}
    \vertex (o) at (0,-0.1) {};
    \vertex (d1) at (2,0) {};
    \vertex (v1) at (0.3,0) {};
    \vertex (v2) at (-0.3,0) {};
    \vertex (d2) at (-2,0) {};
    \vertex (b) at (0,2.5);
    \vertex (A0-v) at (0,0.3) {};
    \vertex (A0) at (0,2) {};
    \node at (0.4,1.8) {$A_0^{a}$};
    \node at (-2.2,0) {$h$};
    \node at (1.2,-0.4) {$p + \frac{k}{2}$};
    \node at (-1.2,-0.4) {$p - \frac{k}{2}$};
    \propag [double, with arrow=0.5] (v1) to (d1);
    \propag [double, with arrow=0.5] (d2) to (v2);
    \propag [bos, mom'={[arrow shorten=0.35] $\k$}] (A0) to (A0-v);
    \node[grayblob, scale=1.5](0,0.5) at (0,0) {};
   \end{feynhand}
 \end{tikzpicture}}
 }
 \,.
 \label{eq:form-factor-diagram}
\end{equation}
Accordingly, we evaluate the Feynman diagram depicted in Eq.~\eqref{eq:form-factor-diagram} with the following kinematic setup:
\begin{equation}\label{kinematic-point}
  k = (0, \k),\quad
  p = \biggl( -B + \frac{\k^2}{8m_h}, \bm{0} \biggr),
\end{equation}
where the ``photon" is assumed to couple to either the electric charge or dineutron densities, as we now explain.

In what follows, we consider two distinct ``photons": one is the usual photon $A_\mu^{\mathrm{em}}$ that couples to the electric charge, and the other is a hypothetical one $A_\mu^{\mathrm{nn}}$---referred to as the ``dineutron photon''---which minimally couples to the neutrons in the halo but not to those in the core.
The dineutron photon can be theoretically introduced as a U(1) gauge boson coupled to $B - \frac{A}{Z} Q$, where $B$ is the baryon number, $Q$ is the electric charge, and $Z$ and $A$ are the atomic and mass numbers of the core, respectively.
Recalling the charges assigned to each field, we introduce terms responsible for the two distinct photon couplings by replacing the partial derivatives of each field in the effective Lagrangian \eqref{eq:bare-Lag}--\eqref{eq:neutron-Lag} with the following covariant derivatives:
\begin{subequations}\label{eq:gauge-coupling}
\begin{align}
 D_\mu h 
 &\equiv (\partial_\mu - \rmi Z A_\mu^{\mathrm{em}} - 2 \rmi A_\mu^{\mathrm{nn}} ) h\,, 
 \\
 D_\mu \phi 
 &\equiv (\partial_\mu - \rmi Z A_\mu^{\mathrm{em}}) \phi\,,
 \\
 D_\mu n
 &\equiv (\partial_\mu - \rmi A_\mu^{\mathrm{nn}}) n\,,
 \label{eq:covariant-derivataive-n}
 \\
 D_\mu d
 &\equiv (\partial_\mu - 2 \rmi A_\mu^{\mathrm{nn}}) d\,,
 \label{eq:covariant-derivataive-d}
\end{align}    
\end{subequations}
where $\mu$ runs over the spacetime indices $t,x,y,z$.
In this framework, the charge (or dineutron) form factor can be computed by assigning these ``photon" in Eq.~\eqref{eq:form-factor-diagram} to be either the usual photon or the dineutron photon.

In addition to the two distinct photons, one can introduce a mass ``photon" to calculate the matter radius $\langle r_m^2 \rangle$. 
However, the matter radius can also be derived from the charge and dineutron radii, as they are interrelated.
To demonstrate this, recall that our EFT models the core and neutrons as point-like particles. 
Let $\r_{\mathrm{core}}$ represent the core's coordinates, and $\r_1$ and $\r_2$ represent the coordinates of two neutrons.
For convenience, we define the Jacobi coordinates as follows:
\begin{align}
  \r &= \r_1 - \r_2, 
  \label{eq:inter-neutron-dis}
  \\
  \bm{\rho} &= \sqrt{\frac A{A+2}} (\r_1 + \r_2 - 2\r_{\mathrm{core}}),
  \label{eq:Jacobi-rho}
  \\
  \bm{R}_\text{cm} & = \frac1{A+2}(\r_1+\r_2 + A\r_{\mathrm{core}}).
\end{align}
Here, the normalization of $\bm{\rho}$ is chosen so that the kinetic part of the first-quantized Hamiltonian becomes symmetric under rotations in the six-dimensional space $(\r, \bm\rho)$.
\begin{equation}
  -\frac12 \left( \frac{\d^2}{\d\r_1^2} + \frac{\d^2}{\d\r_2^2} + \frac1A\frac{\d^2}{\d\r_{\mathrm{core}}^2}\right) = -\frac{\d^2}{\d\r^2} -\frac{\d^2}{\d\bm{\rho}^2} - \frac1{2(A+2)}\frac{\d}{\d\bm{R}_\text{cm}^2} \,,
\end{equation}
We note that the ground state of the halo nucleus is assumed to be symmetric under the exchange of the two neutrons $\r_1 \leftrightarrow \r_2$, which implies $\< \r \cdot \bm{\rho} \> =0$.
The charge, dineutron, and matter radii are then expressed as
\begin{subequations}\label{eq:jacobicoordinates}
\begin{align}
 \<r_c^2\> &= \< (\r_{\mathrm{core}}-\bm{R}_\text{cm})^2\> = 
 \frac{\< \bm{\rho}^2\>}{A(A+2)}\,,
 \\
 \<r_n^2\> 
 &= \frac{1}{2}
 \Big[
  \< (\r_1-\bm{R}_\text{cm})^2\>  + \< (\r_2-\bm{R}_\text{cm})^2\> 
 \Big]
 = \frac{1}{4} 
 \left[ 
  \frac{A}{A+2} \< \bm{\rho}^2\> + \<\r^2\>
 \right]
 \,,
 \\
 \<r_m^2\> &= \frac1{A+2} 
 \Big[
 \< (\r_1-\bm{R}_\text{cm})^2 \>
 + \<(\r_2-\bm{R}_\text{cm})^2\> 
 + A \<(\r_{\mathrm{core}}-\bm{R}_\text{cm})^2\>
 \Big]
 = \frac{\<\bm{\rho}^2\> + \<\bm{r}^2\>}{2(A+2)}\,.
  \label{eq:jacobicoordinates-rm}
\end{align}
\end{subequations}
As a result, the matter radius can be expressed in terms of the dineutron and charge radii as
\begin{equation}\label{eq:radius-relations}
 \<r_m^2\> = \frac{2}{A+2} \<r_n^2\> + \frac{A}{A+2} \<r_c^2\>
 \,.
\end{equation}
Moreover, the ratio of the matter radius to the charge radius is directly related to the ratio of the mean squares of the two Jacobi coordinates:
\begin{equation}\label{eq:ratio-matter-charge}
  \frac{\<r_m^2\>}{\<r_c^2\>} = \frac A2 \left( 1+ \frac{\<\r^2\>}{\<\bm{\rho}^2\>}\right).
\end{equation}

\subsection{Charge radius}
\label{sec:Charge-radius}

Let us now compute the effective-range correction to the charge radius. 
The calculation follows the same procedure outlined in the Supplemental Material of Ref.~\cite{Hongo:2022sdr}. 
For clarity and readability, we briefly summarize the calculation here.

\begin{figure}[t]
 \centering
 \includegraphics[width=0.38\textwidth]{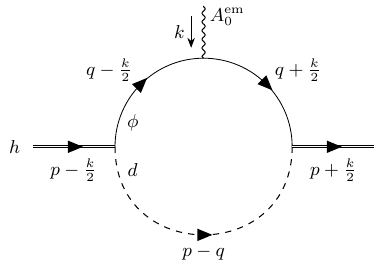}
 \cout{
  \scalebox{1.0}{
  \begin{tikzpicture}[baseline=(o.base)]
   \begin{feynhand}
    \vertex (d2) at (3,0) {};
    \vertex (v2) at (1.5,0);
    \vertex (d1) at (-3,0) {};
    \vertex (v1) at (-1.5,0);
    \vertex (A0-v) at (0,1.5);
    \vertex (A0) at (0,2.5) {};
    \node at (-3.2,0) {$h$};
    \node at (-1.2,0.4) {$\phi$};
    \node at (-1.2,-0.4) {$d$};
    \node at (0.4,2.2) {$A_0^{\mathrm{em}}$};
    \node at (2.2,-0.4) {$p + \frac{k}{2}$};
    \node at (-2.2,-0.4) {$p - \frac{k}{2}$};
    \node at (0,-1.8) {$p-q$};
    \node at (-1.6,1.3) {$q-\frac{k}{2}$};
    \node at (1.6,1.3) {$q+\frac{k}{2}$};
    \propag [double, with arrow=0.5] (d1) to (v1);
    \propag [with arrow=0.5] 
    (v1) to [out=90,in=180] (A0-v);
    \propag [with arrow=0.5] (A0-v) to [out=0,in=90] (v2);
    \propag [sca, with arrow=0.5] (v1) to [out=270,in=270, looseness=1.7] (v2);
    \propag [double, with arrow=0.5] (v2) to (d2);
    \propag [bos, mom'={$k$}] (A0) to (A0-v);
   \end{feynhand}
  \end{tikzpicture}}
  }
  \caption{The Feynman diagram that determines the charge form factor of
  the halo nucleus.}
\label{fig:charge-ff-pqk}
\end{figure}
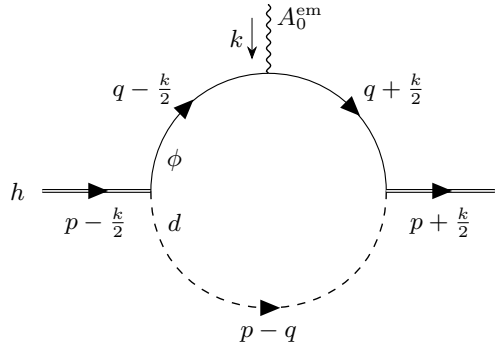

As we explained in the previous section, one can read off the charge radius from the $\k^2$-coefficient of the charge form factor $F_c(\k)$.
It is diagrammatically given by Eq.~\eqref{eq:form-factor-diagram} with the usual photon and on-shell halo satisfying the kinematics in Eq.~\eqref{kinematic-point}.
Identifying the blob of Eq.~\eqref{eq:form-factor-diagram} as the sum of a direct coupling and the loop contribution $\Gamma (k,p)$ shown in Fig.~\ref{fig:charge-ff-pqk}, we find the charge form factor is given by 
\begin{equation}
 F_c (\k) = Q_c \big[ Z_h + \Gamma_c (k,p) \big]\,.
\end{equation}
with the field renormalization $Z_h$ given in Eq.~\eqref{eq:Zh} (recall that $Q_c = Z$ is the atomic number of the core).
The loop contribution is readily evaluated as 
\begin{align}
 \Gamma_c (k,p) 
 &= (\rmi g)^2 \int \frac{\diff^4 q}{(2\pi)^4} 
 \rmi G_\phi \left( q - \frac{k}{2} \right)
 \rmi G_\phi \left( q + \frac{k}{2} \right)
 \rmi D (p-q)
 \nonumber \\
 &= g^2 \int \frac{\diff \q}{(2\pi)^3} 
 \frac{1}{\vep_{\q - \frac{\k}{2}} - \vep_{\q + \frac{\k}{2}}}
 \left[ 
  D \left( -B + \frac{\k^2}{8m_h} - \vep_{\q - \frac{\k}{2}}, - \q \right)
  D \left( -B + \frac{\k^2}{8m_h} - \vep_{\q + \frac{\k}{2}}, - \q \right)
 \right],
\end{align}
where we performed the $q_0$-integral by closing the lower-half contour with $\vep_{\q} = \q^2/(2m_\phi)$ and used the kinematics~\eqref{kinematic-point} to obtain the second line.
Substituting our parametrization of the dimer propagator in Eqs.~\eqref{eq:D-f}--\eqref{eq:def-f}, we obtain the charge form factor as
\begin{equation}
 \frac{F_c (\k)}{Q_c}
 = Z_h + 4\pi g^2 \int \frac{\diff \q}{(2\pi)^3} \frac{m_\phi}{\q \cdot \k}
  \left[
   f \left( B_{\q} + \frac{\k^2}{4m_hm_\phi} - \frac{\q \cdot \k}{2m_\phi}; a, r_0 \right)
   - f \left( B_{\q} + \frac{\k^2}{4m_hm_\phi} + \frac{\q \cdot \k}{2m_\phi}; a, r_0 \right)
  \right],
\end{equation}
with $B_\q$ defined in Eq.~\eqref{eq:Bq}.

In the $\k = 0$ limit, the right-hand side of this equation reduces to one of our renormalization conditions, Eq.~\eqref{eq:Zh}, reproducing the expected result $F_c (\k = 0) = Q_c$.
Expanding the equation up to $O(\k^2)$, we extract the charge radius from its coefficient as
\begin{equation}
  \< r_c^2 \> = \frac{4\pi g^2}{m_\phi m_h} \! \int\!
  \frac{\diff \q}{(2\pi)^3}\left[
  \frac 32 f''(B_\q; a,r_0)
  + \frac{\q^2}{6\mu} f'''(B_\q; a, r_0) \right],
  \label{eq:chargeradius}
\end{equation}
where $f^\prime(x;a,r_0)$ denotes the derivative of $f(x;a,r_0)$ with respect to the first argument $x$
\footnote{
We note that this result matches the expression in Eq.~(S24) of Ref.~\cite{Hongo:2022sdr}, with the function $f(x;a)$, defined in Eq.~\eqref{eq:def-fa} (denoted as $f_a(x)$ in Ref.~\cite{Hongo:2022sdr}), replaced by $f(x;a,r_0)$, as defined in Eq.~\eqref{eq:def-f}.
}.

We can analytically evaluate the remaining $\q$-integral in Eq.~\eqref{eq:chargeradius}. 
Applying the expansion \eqref{eq:def-f}, we obtain the charge radius with the first-order effective-range correction as (see Appendix \ref{ap:Charge-radius} for details)
\begin{equation}
  \langle r_c^2\rangle=\frac{4g^2}{\pi B}\frac{A^{1/2}}{(A+2)^{5/2}}
  \left[
   f_c(\beta) 
   + \tilde{r}_0 f_c^{(1)}(\beta)+O(\tilde{r}_0^2)
  \right].
  \label{eq:Charge-radius}
\end{equation}
Here, we introduced two dimensionless functions $f_c(\beta)$ and $f_c^{(1)}(\beta)$ as
\begin{align}
\label{eq:fc0}
  f_c(\beta) = -\!\int\limits_1^\infty\! \diff y\, \frac{\partial_y f(y;-1/\beta)}{\sqrt{y-1}} 
  &= \begin{cases}
    \displaystyle{\frac1{1-\beta^2}} - \displaystyle{\frac{\beta\arccos\beta}{(1-\beta^2)^{3/2}}} \,, &     \beta<1,\vspace{6pt} \\
    \displaystyle{\frac13}\,, &\beta =1, \vspace{3pt} \\
    \displaystyle{-\frac1{\beta^2-1}} + \displaystyle{\frac{\beta\arccosh\beta}{(\beta^2-1)^{3/2}}}\,, &\beta>1,
    \end{cases}
 \\
\label{eq:fc1}
  f_c^{(1)}(\beta) 
  = - \frac 12 \!\int\limits_1^\infty\! \diff y\,
  \frac{ \d_y [y f^2(y;-1/\beta)]}{\sqrt{y-1}} 
  &= \begin{cases}
  \displaystyle{-\frac\beta2 \left[ \frac{2+\beta^2}{(1-\beta^2)^2} - \frac{3\beta\arccos\beta}{(1-\beta^2)^{5/2}} \right]},  & \beta <1,\vspace{6pt}\\
  -\displaystyle{\frac15}\,,& \beta = 1,\vspace{6pt} \\
  \displaystyle{-\frac\beta2 \left[ \frac{2+\beta^2}{(\beta^2-1)^2} - \frac{3\beta\arccosh\beta}{(\beta^2-1)^{5/2}} \right]},  & \beta >1,
  \end{cases}
\end{align}
in terms of two dimensionless parameters
\begin{equation}
\label{betarho}
\beta = - \frac{1}{a\sqrt B}\equiv \sqrt{\frac{\epsilon_n}B}\,, \qquad
\tilde{r}_0=r_0\sqrt{B}=\sqrt{\frac B{\epsilon_0}},
\end{equation}
with $\epsilon_n \equiv \hbar^2/(m_n a^2) \approx 0.12\,\text{MeV}$ and
$\epsilon_0\equiv\hbar^2/(m_n r_0^2)\approx 5.5\,\text{MeV}$.
Note that the functions $f_c(\beta)$ and $f_c^{(1)}(\beta)$ are continuous across $\beta=1$. (To highlight their real-valued nature, we present alternative expressions for different parameter ranges. 
We give here also the asymptotic behaviors of these functions for the small and large $\beta$ limits as
\begin{equation}\label{eq:fc-asymptotics}
    f_c(\beta) \to \begin{cases}
    1 + O(\beta), 
    & \beta\ll 1, \vspace{6pt}\\
    \displaystyle{\frac{\ln(2\beta)-1}{\beta^2} + O\left(\frac{\ln\beta}{\beta^4}\right)}, & \beta\gg 1,
    \end{cases}  \qquad f_c^{(1)}(\beta) \to \begin{cases}
    -\beta + O(\beta^2) & \beta\ll 1\vspace{6pt} \\
    \displaystyle{-\frac{1}{2\beta} - \frac{3\ln(2\beta)-4}{2\beta^3} + O\left(\frac{\ln\beta}{\beta^5}\right)}& \beta\gg 1
    \end{cases}
\end{equation}
These limits describe the asymptotic behavior of the charge radius when the two-neutron separation energy $B$ is either much larger ($\beta \ll 1$)  or much smaller ($\beta \gg 1$) than the two-neutron virtual energy $\epsilon_n\approx0.12$~MeV.

\subsection{Dineutron and matter radius}
\label{sec:Neutron-Matter-radius}

We next evaluate the dineutron radius and matter radius of the halo nuclei, which demonstrates a peculiar halo structure.
As the intermediate step toward the matter radius, we first compute the dineutron radius of the halo nuclei $\<r_n^2\>$.

\begin{figure}[t]
 \centering
 \includegraphics[width=0.38\textwidth]{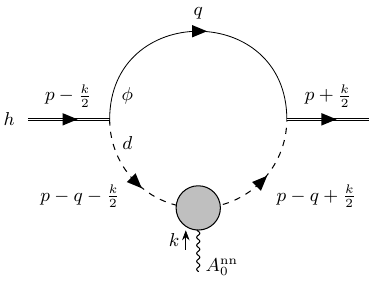}
 \cout{
  \scalebox{1.0}{
  \begin{tikzpicture}[baseline=(o.base)]
   \begin{feynhand}
    \vertex (d2) at (3,0) {};
    \vertex (v2) at (1.5,0);
    \vertex (d1) at (-3,0) {};
    \vertex (v1) at (-1.5,0);
    \vertex (A0-v) at (0,-1.5);
    \vertex (A0) at (0,-2.7) {};
    \node at (-3.2,0) {$h$};
    \node at (-1.2,0.4) {$\phi$};
    \node at (-1.2,-0.4) {$d$};
    \node at (0.4,-2.5) {$A_0^{\mathrm{nn}}$};
    \node at (2.2,0.4) {$p + \frac{k}{2}$};
    \node at (-2.2,0.4) {$p - \frac{k}{2}$};
    \node at (0,1.8) {$q$};
    \node at (-2,-1.3) {$p-q-\frac{k}{2}$};
    \node at (2,-1.3) {$p-q+\frac{k}{2}$};
    \propag [double, with arrow=0.5] (d1) to (v1);
    \propag [with arrow=0.5] (v1) to [out=90,in=90, looseness=1.7] (v2);
    \propag [sca, with arrow=0.5] (v1) to [out=270,in=180] (A0-v);
    \propag [sca, with arrow=0.5] (A0-v) to [out=0,in=270] (v2);
    \propag [double, with arrow=0.5] (v2) to (d2);
    \propag [bos, mom={[arrow shorten=0.35] $k$}] (A0) to (A0-v);
    \node[grayblob, scale=1.0] at (0,-1.5) {};
   \end{feynhand}
  \end{tikzpicture}}
  }
  \caption{The Feynman diagram that determines the dineutron form factor of
  the halo nucleus.}
 \label{fig:neutron-form-factor}
\end{figure}
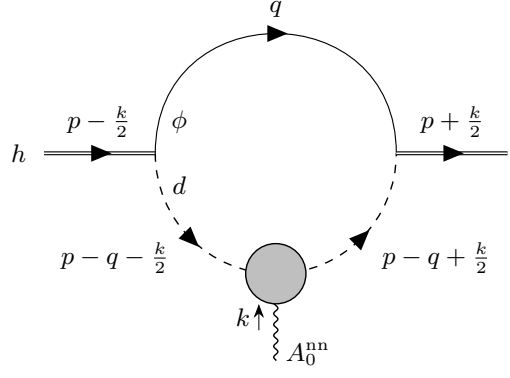

As we explained in Sec.~\ref{sec:Geometry}, the dineutron radius is computed from the corresponding form factor attached to the dineutron photon $A_\mu^{\mathrm{nn}}$.
Noting that the dineutron photon couples to both neutron and dimer fields [recall Eqs.~\eqref{eq:covariant-derivataive-n} and \eqref{eq:covariant-derivataive-d}], it is useful to define the effective vertex $\Gamma_{dd\gamma}(k,p)$ as
\vspace{30pt}
\begin{equation}
 \Gamma_{dd\gamma} (k,p) :=
  \parbox{4.0cm}{\vspace{-34pt}\includegraphics[width=4.0cm]{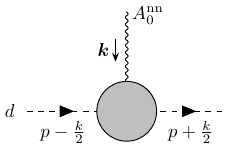}}
 \cout{
 \scalebox{0.9}{
  \begin{tikzpicture}[baseline=(o.base)]
   \begin{feynhand}
    \vertex (o) at (0,-0.1) {};
    \vertex (d1) at (2,0) {};
    \vertex (v1) at (0.3,0) {};
    \vertex (v2) at (-0.3,0) {};
    \vertex (d2) at (-2,0) {};
    \vertex (b) at (0,2);
    \vertex (A0-v) at (0,0.3) {};
    \vertex (A0) at (0,2) {};
    \node at (0.4,1.8) {$A_0^{\mathrm{nn}}$};
    \node at (-2.2,0) {$d$};
    \node at (1.2,-0.4) {$p + \frac{k}{2}$};
    \node at (-1.2,-0.4) {$p - \frac{k}{2}$};
    \propag [chasca, with arrow=0.5] (v1) to (d1);
    \propag [chasca, with arrow=0.5] (d2) to (v2);
    \propag [bos, mom'={[arrow shorten=0.35] $\k$}] (A0) to (A0-v);
    \node[grayblob, scale=1.5](0,0.5) at (0,0) {};
   \end{feynhand}
 \end{tikzpicture}}
 }
 \label{eq:effective-vertex-ddgamma}
\end{equation}
which appears as the subgraph for the dineutron form factor.
With this effective vertex, we find that the dineutron form factor is given by the sum of the direct coupling and loop contribution $\Gamma_n (k,p)$ shown in Fig.~\ref{fig:neutron-form-factor} as 
\begin{equation}\label{eq:Fnk}
 \begin{split}
  F_n(\k) 
  &= 2 Z_h + (\rmi g)^2 \!\int\!\frac{\diff^4 q}{(2\pi)^4}\,
  \rmi G_\phi (q) \rmi D\left(p-q-\frac k2\right) \Gamma_{dd\gamma}(k,p-q)
  \rmi D \left(p-q+\frac k2\right) 
  \\
  &= 2 Z_h + g^2 \!\int\!\frac{\diff \q}{(2\pi)^3}\,
  D\left(p-q-\frac k2\right)\Gamma_{dd\gamma}(k,p-q)
  D\left(p-q+\frac k2\right) \biggl|_{q_0=\frac{\q^2}{2m_\phi}}  \,,
 \end{split}
\end{equation}
where we performed the $q_0$ integral by completing the contour in the lower half-plane [$k$ and $p$ are again set at the kinematic point as in Eq.~\eqref{kinematic-point}].
To proceed, we need to evaluate $\Gamma_{dd\gamma} (k,p)$ up to $O(\k^2)$ to obtain the dineutron radius.

\begin{figure*}[t]
 \centering
 \includegraphics[width=0.75\textwidth]{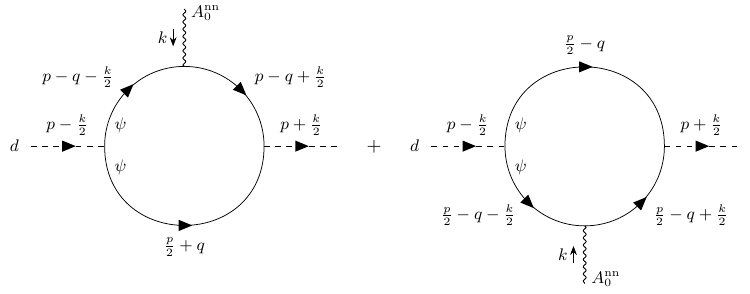}
 \cout{
  \scalebox{0.9}{
  \begin{tikzpicture}[baseline=(o.base)]
   \begin{feynhand}
    \vertex (d2) at (3,0) {};
    \vertex (v2) at (1.5,0);
    \vertex (d1) at (-3,0) {};
    \vertex (v1) at (-1.5,0);
    \vertex (A0-v) at (0,1.5);
    \vertex (A0) at (0,2.7) {};
    \node at (-3.2,0) {$d$};
    \node at (-1.2,0.4) {$\psi$};
    \node at (-1.2,-0.4) {$\psi$};
    \node at (0.4,2.5) {$A_0^{\mathrm{nn}}$};
    \node at (2.2,0.4) {$p + \frac{k}{2}$};
    \node at (-2.2,0.4) {$p - \frac{k}{2}$};
    \node at (0,-1.9) {$\frac{p}{2}+q$};
    \node at (-2,1.3) {$p-q-\frac{k}{2}$};
    \node at (2,1.3) {$p-q+\frac{k}{2}$};
    \propag [sca, with arrow=0.5] (d1) to (v1);
    \propag [with arrow=0.5] (v1) to [out=270,in=270, looseness=1.7] (v2);
    \propag [with arrow=0.5] (v1) to [out=90,in=180] (A0-v);
    \propag [with arrow=0.5] (A0-v) to [out=0,in=90] (v2);
    \propag [sca, with arrow=0.5] (v2) to (d2);
    \propag [bos, mom'={[arrow shorten=0.35] $k$}] (A0) to (A0-v);
   \end{feynhand}
  \end{tikzpicture}}
  $~+~$
  \scalebox{0.9}{
  \begin{tikzpicture}[baseline=(o.base)]
   \begin{feynhand}
    \vertex (d2) at (3,0) {};
    \vertex (v2) at (1.5,0);
    \vertex (d1) at (-3,0) {};
    \vertex (v1) at (-1.5,0);
    \vertex (A0-v) at (0,-1.5);
    \vertex (A0) at (0,-2.7) {};
    \node at (-3.2,0) {$d$};
    \node at (-1.2,0.4) {$\psi$};
    \node at (-1.2,-0.4) {$\psi$};
    \node at (0.4,-2.5) {$A_0^{\mathrm{nn}}$};
    \node at (2.2,0.4) {$p + \frac{k}{2}$};
    \node at (-2.2,0.4) {$p - \frac{k}{2}$};
    \node at (0,1.9) {$\frac{p}{2}-q$};
    \node at (-2,-1.3) {$\frac{p}{2}-q-\frac{k}{2}$};
    \node at (2,-1.3) {$\frac{p}{2}-q+\frac{k}{2}$};
    \propag [sca, with arrow=0.5] (d1) to (v1);
    \propag [with arrow=0.5] (v1) to [out=90,in=90, looseness=1.7] (v2);
    \propag [with arrow=0.5] (v1) to [out=270,in=180] (A0-v);
    \propag [with arrow=0.5] (A0-v) to [out=0,in=270] (v2);
    \propag [sca, with arrow=0.5] (v2) to (d2);
    \propag [bos, mom={[arrow shorten=0.35] $k$}] (A0) to (A0-v);
   \end{feynhand}
  \end{tikzpicture}}
  }
  \caption{The diagrams contributing to the effective coupling of the dineutron photon to the dimer field.}
\label{fig:effective-vertex}
\end{figure*}
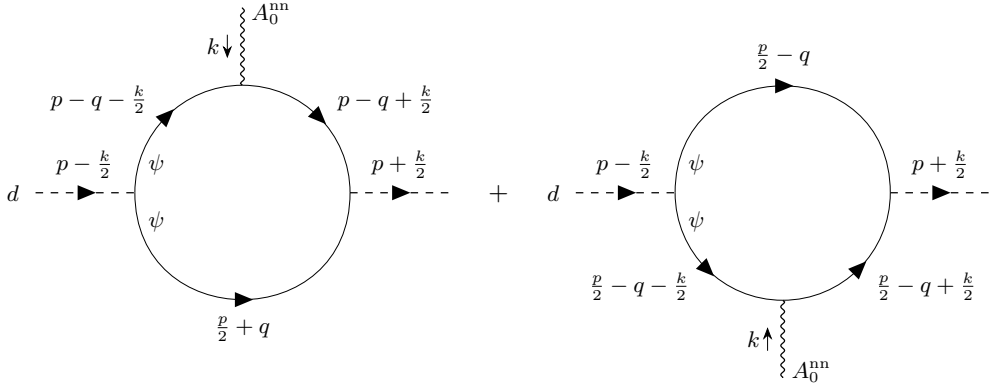

The effective coupling of the dimer to the dineutron photon, $\Gamma_{dd\gamma}(k,p)$, at zeroth order in $r_0$ is given by the sum of the two Feynman diagrams shown in Fig.~\ref{fig:effective-vertex}, which was computed in Ref.~\cite{Hongo:2022sdr}.
The effective-range correction in the Lagrangian \eqref{eq:neutron-Lag} introduces a contact-term correction to the effective coupling, leading to
\begin{equation}\label{eq:neutron-dimer-vertex}
 \begin{split}
  \Gamma_{dd\gamma}(k,p) 
  &= \left[ 
  \rmi^2 \int \frac{\diff^4 q}{(2\pi)^4} 
  \rmi G_\psi \left( \frac{p}{2} - q - \frac{k}{2} \right)
  \rmi G_\psi \left( \frac{p}{2} - q + \frac{k}{2} \right)
  \rmi G_\psi \left( \frac{p}{2} + q \right)
 + (q \to -q) 
 \right]
 + \frac{1}{4\pi} r_0
  \\
  &\simeq \frac1{4\pi}\left[
  \frac{1}{\sqrt{-P_0}}
  -\frac 5{96} \frac{ \k^2}{(-P_0)^{3/2}}
  + \frac 1{32} \frac{K_0^2}{(-P_0)^{5/2}} - r_0 \right] ,
 \end{split}
\end{equation}
where we performed the $q$-integration and retained terms up to $O(\k^2)$ in the second line with two Galilean-invariant combinations $P_0=p_0-\p^2/4$, $K_0=k_0-\frac12\p\cdot\k$
\footnote{
We refer to Supplemental Material of Ref.~\cite{Hongo:2022sdr} for the detailed procedure to derive the first three terms.
}.

Substituting Eq.~(\ref{eq:neutron-dimer-vertex}) into Eq.~(\ref{eq:Fnk}) and
expanding to order $\k^2$, one finds the dineutron radius
\begin{equation}\label{eq:rn}
  \begin{split}
    \< r_n^2\> = &-\pi g^2 \!\int\!\frac{\diff \q}{(2\pi)^3}\,
  \Bigg\{ 
  \left[ \left( -\frac5{8} + \frac3{4m_h}\right)\frac1{B_\q^{3/2}} + \frac1{32}\frac{\q^2}{B_\q^{5/2}}
  \right]f^2(B_\q;a,r_0)\\
  & + \left( \frac1{\sqrt{B_\q}} - r_0 \right)
  \left[ \frac{3m_\phi}{2m_h} f(B_\q;a,r_0)f'(B_\q;a,r_0) + \frac1{4}\q^2 \left(f(B_\q;a,r_0)f''(B_\q;a,r_0)-f'^2(B_\q;a,r_0)\right) \right]
  \Bigg\}.
  \end{split}
\end{equation}

After performing the above integrals, we obtain the analytic result for the dineutron radius with the first-order effective-range correction as (see Appendix \ref{ap:Neutron-Matter-radius} for details)
\begin{equation}
  \< r_n^2 \> = \frac{g^2}{\pi B}
  \left( \frac{A}{A+2} \right)^{3/2}
  \left[ 
   f_n (\beta) + \frac A{A+2} f_c (\beta)
   + \tilde{r}_0  
   \left(
    f_n^{(1)} (\beta) + \frac A{A+2} f_c^{(1)} (\beta)
   \right)
   + O (\tilde{r}_0^2)
  \right].
  \label{eq:Neutron-Matter-radius}
\end{equation}
with the dimensionless parameters defined in Eq.~\eqref{betarho}.
In addition to $f_n (\beta)$ and $f_n^{(1)} (\beta)$ already introduced in Eqs.~\eqref{eq:fc0} and \eqref{eq:fc1}, we here defined two dimensionless functions $f_n (\beta)$ and $f_n^{(1)} (\beta)$ as
\begin{align}
\label{eq:fn0}
  f_n (\beta) = 
  \int\limits_1^\infty \diff y\,
  \frac{\sqrt{y-1}}{2y^{3/2}} \left( \frac1{\sqrt y+\beta}\right)^2 
  &=\begin{cases}
     \displaystyle{\frac1{\beta^3}} \biggl[ \pi-2\beta +
    (\beta^2-2)\displaystyle{
    \frac{\arccos\beta}{\sqrt{1-\beta^2}} }\biggr], & \beta <1,\vspace{4pt}\\
    \pi-3, & \beta =1, \vspace{4pt}\\
    \displaystyle{\frac1{\beta^3}} \biggl[  \pi-2\beta +
    (\beta^2-2)\displaystyle{
    \frac{\arccosh\beta}{\sqrt{\beta^2-1}} }\biggr], & \beta>1,
    \end{cases}
 \\
\label{eq:fn1}
  f_n^{(1)}(\beta) 
  = \int\limits_1^\infty \diff y\,
 \frac{\sqrt{y-1}}{2y^{1/2}} \left( \frac1{\sqrt y+\beta}\right)^3 
  &= \begin{cases}
     \displaystyle{\frac12 \left[ - \frac\beta{1-\beta^2}
     + \frac{\arccos\beta}{(1-\beta^2)^{3/2}}
     \right]}, & \beta < 1, \vspace{6pt}\\
     \displaystyle{\frac13}\,, & \beta=1,\vspace{6pt} \\
     \displaystyle{\frac12 \left[ \frac\beta{\beta^2-1}
     - \frac{\arccosh\beta}{(\beta^2-1)^{3/2}}
     \right]}, & \beta > 1.
  \end{cases}
\end{align}
As with $f_c (\beta)$ and $f_c^{(1)} (\beta)$, the functions $f_n(\beta)$ and $f_n^{(1)}(\beta)$ are continuous across $\beta=1$ (We provide alternative expressions for these functions at $\beta<1$ and $\beta>1$ to explicitly demonstrate that they remain real when their arguments are real).
We also present their asymptotic behaviors in the limits of large and small $\beta$ as follows:
\begin{equation}\label{eq:fn-asymptotics}
    f_n(\beta) \to \begin{cases}
    \displaystyle{\frac13} + O(\beta), 
    & \beta\ll 1, \vspace{6pt}\\
    \displaystyle{\frac{\ln(2\beta)-2}{\beta^2} + O\left(\frac1{\beta^3}\right)}, & \beta\gg 1,
    \end{cases}  \qquad f_n^{(1)}(\beta) \to \begin{cases}
    \displaystyle{\frac\pi4} + O(\beta) & \beta\ll 1,\vspace{6pt} \\
    \displaystyle{\frac{1}{2\beta} + O\left(\frac{\ln\beta}{\beta^3}\right)}& \beta\gg 1.
    \end{cases}
\end{equation}

With the derived charge and dineutron radii, Eq.~\eqref{eq:radius-relations} allows us to extract the matter radius, including the effective-range correction, as
\begin{equation}\label{eq:rm}
  \<r_m^2\> = 
  \frac{2g^2}{\pi B}\frac{A^{3/2}}{(A+2)^{5/2}} 
  \left[
   f_c(\beta) + f_n(\beta)
   + \tilde{r}_0 \big( f_c^{(1)} (\beta) + f_n^{(1)} (\beta) \big)
  \right].
\end{equation}

\subsection{Magnitude of effective-range corrections}
\label{sec:discusion-rms}

The outcome of our calculation is a set of analytic formulas expressing the charge and matter radii of the halo nucleus, given by Eqs.~\eqref{eq:Charge-radius} and \eqref{eq:rm}, respectively, in terms of the coupling constant $g$ and dimensionless functions of two parameters in Eq.~\eqref{betarho}.
In particular, the effective-range corrections---terms proportional to $\tilde{r}_0$---constitute our new result. 
These corrections are given by two dimensionless functions, $f_{c}^{(1)}(\beta)$ and $f_n^{(1)} (\beta)$, defined in Eqs.~\eqref{eq:fc1} and \eqref{eq:fn1}, whose explicit forms appear in the same equations.

Let us now discuss the magnitude of the effective-range corrections to the charge and matter radii.
For this purpose, it is more convenient to work with the mean squares of the Jacobi coordinates.
Recalling Eq.~\eqref{eq:jacobicoordinates}, we find that our results in Eqs.~\eqref{eq:Charge-radius} and \eqref{eq:Neutron-Matter-radius} can be expressed in terms of the rms radii of the Jacobi coordinates as
\begin{subequations}
\begin{align}
  \< \bm{\rho}^2 \> &= 
  \frac{4g^2}{\pi B}\Big(\frac{A}{A+2}\Big)^{3/2}
  \left[
   f_c(\beta) 
   + \tilde{r}_0 f_c^{(1)}(\beta)+O(\tilde{r}_0^2)
  \right],
  \label{eq:rho-expansion} 
  \\
  \< \r^2\> &=\frac{4g^2}{\pi B}\Big(\frac{A}{A+2}\Big)^{3/2}
  \left[
   f_n(\beta) 
   + \tilde{r}_0 f_n^{(1)}(\beta)+O(\tilde{r}_0^2)
  \right],
 \label{eq:r-expansion} 
\end{align}  
\end{subequations}
where both expressions share a common prefactor multiplied by different dimensionless functions.
Note that the charge radius is proportional to $\<\bm{\rho}^2\>$ while the matter radius is proportional to a linear combination of $\<\bm{\rho}^2\>$ and $\<\bm{r}^2\>$ [see Eq.~\eqref{eq:jacobicoordinates}].

\begin{figure}[t]
  \centering
  \begin{minipage}{0.48\textwidth}
      \centering
      \includegraphics[width=\textwidth]{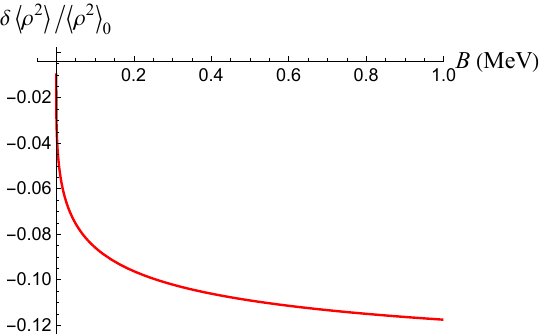}
  \caption{Relative effective-range correction to $\langle\bm{\rho}^2\rangle$, Eq.\ \eqref{eq:rho-relative-correction}, as a function of $B$.}
  \label{fig:rho-relative-correction}
  \end{minipage}\hfill
  \begin{minipage}{0.48\textwidth}
      \centering
      \includegraphics[width=\textwidth]{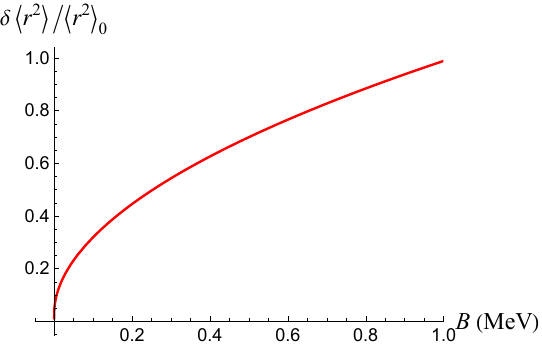}
   \caption{Relative effective-range correction to $\langle\bm{r}^2\rangle$, Eq. \eqref{eq:r-relative-correction}, as a function of $B$.}
  \label{fig:r-relative-correction}
  \end{minipage}
\end{figure}

We now compare the magnitude of the effective-range corrections to the zero-range result.
Denoting the zero-range rms radius for $r$ as $\<r^2\>_0$ and the first-order effective-range correction as $\delta \<r^2\>$, the relative effective-range correction to $\<\bm{\rho}^2\>$ (i.e., the charge radius) is
\begin{equation}
 \frac{\delta \<r_c^2\>}{\<r_c^2\>_0} =
 \frac{\delta\<\bm{\rho}^2\>}{\<\bm{\rho}^2\>_0} = 
 \tilde r_0 \frac{f_c^{(1)}(\beta)}{f_c(\beta)} \,.
 \label{eq:rho-relative-correction}
\end{equation}
From Eq.~\eqref{eq:fc-asymptotics}, the asymptotic behavior at large and small binding energies follows as
\begin{equation}
  \frac{\delta \<r_c^2\>}{\<r_c^2\>_0} 
  = \frac{\delta\<\bm{\rho}^2\>}{\<\bm{\rho}^2\>_0} = \begin{cases} 
  \displaystyle{\frac{r_0}{a}}\,, & B\gg \epsilon_n, \vspace{6pt}\\
  
  \displaystyle{\frac{r_0}{2a} \,\frac1{\ln(2\sqrt{\epsilon_n/B})-1}}\,, & B\ll\epsilon_n .
  \end{cases}
\end{equation}
Note that the correction is negative due to $a<0$ and is suppressed by the factor $r_0/|a|\approx0.15$.
At very small binding energies, an additional logarithmic suppression appears.
We plot this relative correction to the charge radius as a function of $B$ for $B<1$~MeV in Fig. \ref{fig:rho-relative-correction}. 
In all plots, we use $\epsilon_n=0.12$ MeV and $\epsilon_0=5.5$ MeV.

The relative effective-range correction to the Jacobi coordinate $\< \r^2\>$, which describes the inter-neutron distance, is given by
\begin{equation}
  \label{eq:r-relative-correction}
  \frac{\delta\<\bm{r}^2\>}{\<\bm{r}^2\>_0} = 
  \tilde r_0 \frac{f_n^{(1)}(\beta)}{f_n(\beta)} \,.
\end{equation}
Using Eq.~\eqref{eq:fn-asymptotics}, we can determine the asymptotic behavior of the relative correction at large and small binding energies:
\begin{equation}
  \frac{\delta\<\bm{r}^2\>}{\<\bm{r}^2\>_0} 
  = \begin{cases} 
  \displaystyle{ \frac{3\pi}4\sqrt{\frac{B}{\epsilon_0}}}\,, & B\gg \epsilon_n, \vspace{6pt}
  \\
  \displaystyle{\frac12 \frac{r_0}{(-a)}\,\frac1{\ln(2\sqrt{\epsilon_n/B})-2}}\,, & B\ll\epsilon_n .
  \end{cases}
\end{equation}
Figure~\ref{fig:r-relative-correction} shows $\delta\<\r^2\>/\<\r^2\>_0$ as a function of the halo binding energy $B$.
Although the effective-range correction is formally suppressed, Fig.~\ref{fig:r-relative-correction} demonstrates that the correction to $\<\r^2\>$ should be taken into account in practice due to the relatively large numerical factor.
Indeed, the correction to $\<\r^2\>$ remains smaller than the unperturbed value only when
\begin{equation}
  B < \frac{16}{9\pi^2}\epsilon_0 \approx 1~\text{MeV}.  
\end{equation}
Additionally, we note that the correction to $\<\r^2\>$ is relatively larger than the correction to $\<\bm{\rho}^2\>$.

\begin{figure}[t]
    \centering
    \begin{minipage}{0.48\textwidth}
        \centering
        \includegraphics[width=\textwidth]{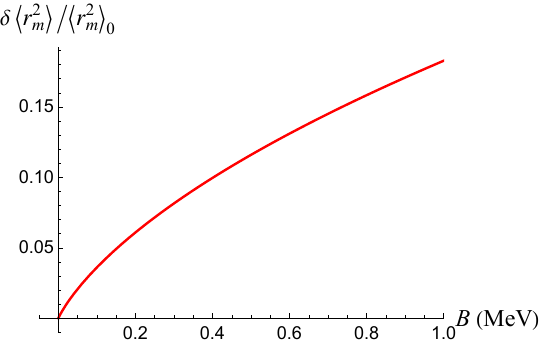}
  \caption{Relative effective-range correction to the mass-radius $\langle r_m^2\rangle$, Eq.\ \eqref{eq:rm-relative-correction}, as a function of $B$.}
\label{fig:rm-relative-correction}
    \end{minipage}\hfill
    \begin{minipage}{0.48\textwidth}
        \centering
        \includegraphics[width=\textwidth]{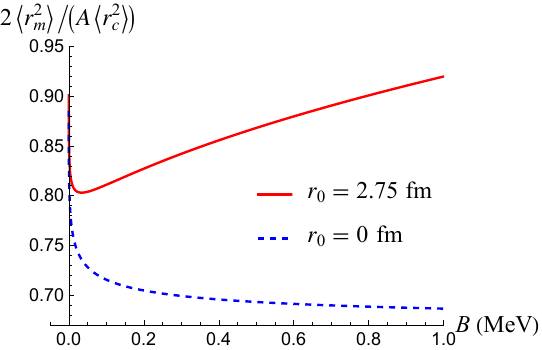}
  \caption{Ratio of the matter radius and the charge radius, Eq.\ \eqref{eq:rm_rc}, with (solid) and without (dashed) the effective correction.}
\label{fig:rm_rc}
    \end{minipage}
\end{figure}

The mean square of the matter radius is proportional to the sum of $\<\bm{\rho^2}\>$ and $\<\bm{r^2}\>$ [recall Eq.~(\ref{eq:jacobicoordinates-rm})].
From Eq.~\eqref{eq:rm}, we immediately obtain the relative correction to the matter adius:
\begin{equation}
 \frac{\delta\<r_m^2\>}{\<r_m^2\>_0} = 
  \tilde r_0 \frac{f_c^{(1)}(\beta)+ f_n^{(1)}(\beta)}{f_c(\beta) + f_n(\beta)} \,,
  \label{eq:rm-relative-correction}
\end{equation}
Using Eqs.~\eqref{eq:fc-asymptotics} and \eqref{eq:fn-asymptotics}, we find that the asymptotic behavior of the relative correction to the matter radius is
\begin{equation}
  \frac{\delta\<r_m^2\>}{\<r_m^2\>_0} = \begin{cases} 
  \displaystyle{ \frac{3\pi}{16}\sqrt{\frac{B}{\epsilon_0}}}\,, & B\gg \epsilon_n, \vspace{6pt}
  \\
  \displaystyle{\frac12 \frac{B}{\sqrt{\epsilon_n\epsilon_0}}}\,, & B\ll\epsilon_n .
  \end{cases}    
\end{equation}
Figure~\ref{fig:rm-relative-correction} shows this correction as a function of $B$. 
From this, we observe that the correction is numerically not so large: it does not exceed approximately 25\% for $B<1$~MeV.  
The largeness of the relative correction to $\<\r^2\>$ is compensated by the fact that this Jacobi coordinate contributes only a small fraction ($1/4$ for $B\gg\epsilon_n$) to the mean square matter radius.

Finally, we discuss the magnitude of the effective-range correction to the ratio of the matter radius to the charge radius.
Substituting Eqs.~\eqref{eq:rho-expansion} and \eqref{eq:r-expansion} into Eq.~\eqref{eq:ratio-matter-charge}, we obtain
\begin{equation}
\label{eq:rm_rc}
  \frac{\<r_m^2\>}{\<r_c^2\>} 
  = \frac A2 \left[ 1 + \frac{f_n(\beta) + \tilde r_0 f_n^{(1)}(\beta))}{f_c(\beta) + \tilde r_0 f_c^{(1)}(\beta )}\right].
\end{equation}
In Fig.~\ref{fig:rm_rc}, we plot this ratio as a function of $B$ for $B<1$ MeV, both with and without the effective-range correction.
As indicated by the range on the vertical axis, the $r_0$ correction increases the ratio $\<r_m^2\>/\<r_c^2\>$ only modestly.
For instance, in the case of $^{22}$C with an assumed binding energy of $B=0.1$~MeV, the correction amounts to approximately 13\%. 
Even for $B=0.975$~MeV, corresponding to the two-neutron separation energy in $^6$He, the correction remains only 34\%.  
However, the correction is large enough to make the dependence of $\<r_m^2\>/\<r_c^2\>$ on $B$ nonmonotonic. 
Additionally, if one expands the ratio in Eq.~\eqref{eq:rm_rc} in $\tilde{r}_0$ and keep only terms up to $O(\tilde r_0)$, the result is indistinguishable from the one without the effective-range correction shown in Fig.~\ref{fig:rm_rc}.

\section{Electromagnetic response}
\label{sec:elemag}

In this section, we investigate the electromagnetic response of halo nuclei.
After analyzing the effective-range correction to the $E1$ dipole strength function in Sec.~\ref{sec:E1-dipole}, we present our results on electric polarizability in Sec.~\ref{sec:electricalpolarizability}.

\subsection{\texorpdfstring{$E1$}{E1} dipole strength function}
\label{sec:E1-dipole}

We here compute the effective-range correction to the $E1$ dipole strength function, defined by 
\begin{equation}\label{eq:def-E1}
 \frac{\diff B(E1)}{\diff\omega} (\omega)
 \equiv \sum_n |\<n|\bm{\mathcal{M}}|0\>|^2 \delta (E_n-E_0-\omega),
\end{equation}
where the dipole operator is defined as $\bm{\mathcal{M}} \equiv \sqrt{\frac{3}{4\pi}} Z e (\r_{\mathrm{core}} - \bm{R}_{\mathrm{cm}} )$, with $\r_{\mathrm{core}}$ and $\bm{R}_{\mathrm{cm}}$ denoting the coordinates of the core and the center of mass, respectively.
We also introduce a sum over all excited states of the halo, $|n\>$, with eigenvalues $E_n$ ($n=0$ corresponding to the ground state).
The $E1$ dipole strength function measures the spectral weight of the excited states induced by the insertion of the dipole operator, characterizing how the halo nucleus responds to external electromagnetic fields.

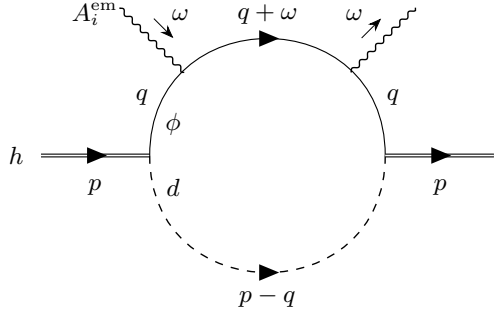
\begin{figure}[t]
 \centering
 \includegraphics[width=0.38\textwidth]{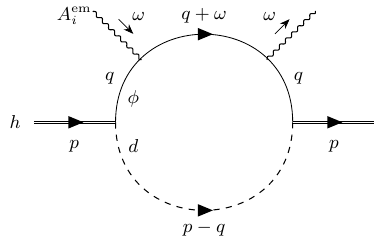}
 \cout{
  \scalebox{1.0}{
  \begin{tikzpicture}[baseline=(o.base)]
   \begin{feynhand}
    \vertex (d2) at (3,0) {};
    \vertex (v2) at (1.5,0);
    \vertex (d1) at (-3,0) {};
    \vertex (v1) at (-1.5,0);
    \vertex (A0-v) at (0,1.5);
    \vertex (Ai) at (-2,2) {};
    \vertex (Ao) at (2,2) {};
    \vertex (Ai-v) at (-1.06,1.06);
    \vertex (Ao-v) at (1.06,1.06);
    \vertex (A0) at (0,2.5) {};
    \node at (-3.2,0) {$h$};
    \node at (-1.2,0.4) {$\phi$};
    \node at (-1.2,-0.4) {$d$};
    \node at (-2.2,1.8) {$A_i^{\mathrm{em}}$};
    \node at (2.2,-0.4) {$p$};
    \node at (-2.2,-0.4) {$p$};
    \node at (0,-1.8) {$p-q$};
    \node at (0,1.8) {$q+\omega$};
    \node at (-1.6,0.75) {$q$};
    \node at (1.6,0.75) {$q$};
    \propag [double, with arrow=0.5] (d1) to (v1);
    \propag [with arrow=0.5] (v1) to [out=90,in=90,looseness=1.7] (v2);
    \propag [sca, with arrow=0.5] (v1) to [out=270,in=270, looseness=1.7] (v2);
    \propag [double, with arrow=0.5] (v2) to (d2);
    \propag [bos, mom={[arrow shorten=0.35] $\omega$}] (Ai) to (Ai-v);
    \propag [bos, mom={[arrow shorten=0.35] $\omega$}] (Ao-v) to (Ao);
   \end{feynhand}
  \end{tikzpicture}}
  }
  \caption{The Feynman diagram determining the $E1$ dipole strength function.}
\label{fig:dis-momentum}
\end{figure}

As shown in Ref.~\cite{Hongo:2022sdr}, by noting that $\partial_t \bm{\mathcal{M}} = \sqrt{\frac{3}{4\pi}} \bm{J}$, where $\bm{J}$ is the total electric current operator carried by the halo, we can express the $E1$ dipole strength function as
\begin{equation}
   \frac{\diff B(E1)}{\diff\omega} (\omega)
   =-\frac{3}{4\pi}\frac1{\pi\omega^2}\Im G_{JJ}(\omega),
\end{equation}
where the two-point correlation function of the total electric current is defined as
\begin{equation}
 \rmi G_{JJ} (\omega) \equiv 
 \int \diff t \rme^{\rmi \omega t} \<0|T\bm{J} (t) \cdot \bm{J} (0)|0\>,
\end{equation}
Since the insertion of the total current operator corresponds to an external photon with vanishing momentum, one finds a diagrammatic representation of $G_{JJ} (\omega)$ in Fig.~\ref{fig:dis-momentum}, where we use the simplified notation $\omega = (\omega,\bm{0})$.
Moreover, we assume that the initial state of the halo nucleus is a static on-shell ground state, and therefore set $p = (-B,\bm{0})$.

By collecting another diagram generated by reversing the direction of momentum flow $\omega = (\omega,\bm{0})$ on the two photon lines in Fig.~\ref{fig:dis-momentum}, we obtain $\rmi G_{JJ}(\omega)$ as
\begin{equation}
 \begin{split}
  \rmi G_{JJ}(\omega) 
  &= (\rmi g)^2 \int \frac{\diff^4 q}{(2\pi)^4} 
  \rmi G_\phi (q) \frac{Ze\q}{m_\phi} \rmi G_\phi (q+\omega) \frac{Ze\q}{m_\phi} \rmi G_\phi (q) 
  \rmi D (p-q)
  \\
  &= \rmi (Ze)^2 \frac{g^2}{m_\phi^2 \omega^2} \int \frac{\diff \q}{(2\pi)^3} q^2 
  D \left(\omega - B - \frac{\q^2}{2m_\phi},-\q \right)
  + (\omega \to - \omega),
 \end{split}
\end{equation}
where we performed the $q_0$-integral by completing the contour in the lower-half plane.
Taking the imaginary part and setting $\omega>0$, we find that the second term vanishes, leading to
\begin{equation}\label{eq:E1-fcn}
 \begin{split}
  \Im G_{JJ}(\omega) 
  &= (Ze)^2 \frac{g^2}{m_\phi^2\omega^2}\!
  \int\! \frac{\diff \q}{(2\pi)^3}\, \q^2
  \Im D\left( \omega-B-\frac{\q^2}{2m_\phi}\,, -\q\right),
  \\
  &= - (Ze)^2 \frac{4\pi g^2}{m_\phi^2\omega^2}\!
  \int\!\frac{\diff \q}{(2\pi)^3}\, \q^2
  \frac{\sqrt{\omega-B-\frac{\q^2}{2\mu}}}
  {\omega-B-\frac{\q^2}{2\mu} + \left[\frac1a -\frac{r_0}2
  \left(\omega-B-\frac{\q^2}{2\mu}\right) \right]^2} \theta\left(\omega-B- \frac{\q^2}{2\mu}\right).
 \end{split}
\end{equation}
which generalizes the result of Ref.~\cite{Hongo:2022sdr} by incorporating the dimer propagator with the effective-range correction.

Performing the remaining $\q$-integral, we obtain the analytic formula for the $E1$ dipole strength with the first-order effective-range correction (see Appendix \ref{ap:dipole-strength-corrected} for details):
\begin{equation}\label{dipole-strength-corrected}
  \frac{\diff B(E1)}{\diff\omega} = \frac3{4\pi}(Ze)^2
  \frac{12g^2}{\pi} \frac{A^{1/2}}{(A+2)^{5/2}}\theta(\omega-B)
  \frac{(\omega-B)^2}{\omega^4}
  \left[ f_{E1} \left( \frac1{-a\sqrt{\omega-B}} \right)
  + \frac{r_0}{a}f^{(1)}_{E1}\left(\frac1{-a\sqrt{\omega-B}}\right)\right],
\end{equation}
where we defined two dimensionless functions
\begin{align}
   f_{E1}(x) &=1-\frac{8}{3}x(1+x^2)^{3/2}+4x^2\Big(1+\frac{2}{3}x^2\Big),
   \label{fE1}
   \\
   f_{E1}^{(1)}(x) &= 1 - 4x(1+2x^2)\sqrt{1+x^2} + 8 x^2 (1+x^2).
   \label{fE11}
\end{align}
The asymptotic behavior of the dimensionless functions $f_{E1}(x)$ and $f_{E1}^{(1)}(x)$ in the limits of small and large $x$ are
\begin{equation}\label{eq:fE1-asymptotics}
    f_{E1}(x) \to \begin{cases}
    \displaystyle{1 - \frac83 x + O(x^2)},  
    & x \ll 1, \vspace{6pt}\\
    \displaystyle{\frac1{6x^2} + O\left(\frac1{x^4}\right)}, & x\gg 1,
    \end{cases}  \qquad f_{E1}^{(1)}(x) \to \begin{cases}
    1 - 4x + O(x^2), & x\ll 1,\vspace{6pt} \\
    \displaystyle{\frac1{16x^4} + O\left(\frac1{x^6}\right)},& x\gg 1.
    \end{cases}
\end{equation}
In particular, in the unitarity limit of the neutron-neutron interaction ($a=\infty$, $r_0=0$), both dimensionless functions reduce to unity, resulting in a particularly simple form for the $E1$ dipole strength:
\begin{equation}\label{eq:dipole-strength-simple}
 \frac{\diff B(E1)}{\diff\omega}\bigg|_\text{unitarity} = \frac3{4\pi}(Ze)^2
 \frac{12g^2}{\pi} \frac{A^{1/2}}{(A+2)^{5/2}}\theta(\omega-B)
 \frac{(\omega-B)^2}{\omega^4} \,.
\end{equation}

It is worth noting that our result, including the effective-range correction, satisfies the sum rule that relates the $E1$ dipole strength function to the charge radius~\cite{Hongo:2022sdr}.
This sum rule follows from the definition of the $E1$ dipole strength in Eq.~\eqref{eq:def-E1} as
\begin{equation}\label{eq:sum-rule}
 \int\limits_B^\infty \!\diff\omega\,\frac{\diff B(E1)}{\diff\omega} 
 = \<0|\bm{\mathcal{M}}^2|0\>
 = \frac{3}{4\pi} (Ze)^2 \<r_c^2\>,
\end{equation}
and serves as a useful self-consistency check of our calculations.
In our calculation, integrating the $E1$ dipole strength over frequency directly confirms that
\begin{equation}\label{eq:integrated-E1}
  \int\limits_B^\infty \!\diff\omega\,\frac{\diff B(E1)}{\diff\omega} =
  \frac3{4\pi}(Ze)^2
  \frac{4g^2}{\pi {B}} \frac{A^{1/2}}{(A+2)^{5/2}}
  \left[ f_c(\beta) + \tilde r_0 f_c^{(1)}(\beta) \right]
  = \frac3{4\pi}(Ze)^2 \< r_c^2\>,
\end{equation}
where we used the following integrals:
\begin{align}
  \int\limits_1^\infty\!\diff y\, \frac{(y-1)^2}{y^4}
  f_{E1}\left( \frac\beta{\sqrt{y-1}}\right) &= \frac13 f_c(\beta),\\
  \int\limits_1^\infty\!\diff y\, \frac{(y-1)^2}{y^4}
  f^{(1)}_{E1}\left( \frac\beta{\sqrt{y-1}}\right) &=
  - \frac1{3\beta} f_c^{(1)}(\beta).
\end{align}
with $f_c(\beta)$ and $f_c^{(1)}(\beta)$ defined in Eqs.~\eqref{eq:fc0} and \eqref{eq:fc1}.

\begin{figure}[t]
  \centering
  \begin{minipage}{0.48\textwidth}
      \centering
      \includegraphics[width=\textwidth]{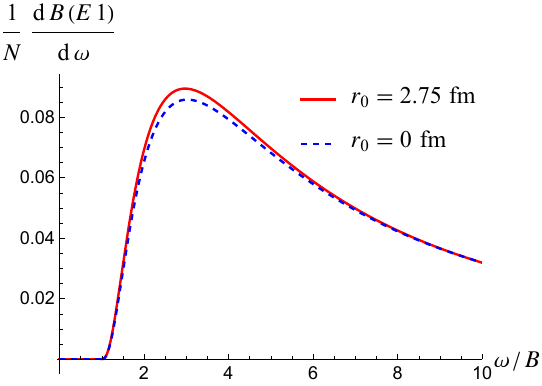}
  \caption{Normalized $E1$ dipole strength, Eq.~\eqref{dipole-strength-normalized}, with (solid) and without (dashed) the effective-range correction for ${}^{22}$C with $B=0.1$ MeV ($\beta=\sqrt{\epsilon_n/B}=1.10$) as a function of $\omega/B$.}
  \label{fig:B01}
  \end{minipage}
  \hfill
  \begin{minipage}{0.48\textwidth}
      \centering
      \includegraphics[width=\textwidth]{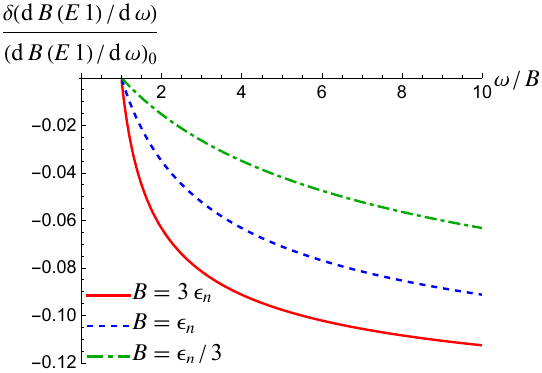}
  \caption{Relative effective-range correction to the $E1$ dipole strength, Eq.\ \eqref{eq:E1-relative-correction}, as a function of $\omega/B$ for $B=3\epsilon_n$, $\epsilon_n$ and $\epsilon_n/3$.
  }
  \label{fig:E1_relative_correction}
  \end{minipage}
\end{figure}

Let us now evaluate the magnitude of the effective-range correction to the $E1$ dipole strength function.
It is important to recall that our calculation does not determine the overall normalization of the $E1$ dipole strength due to the presence of the running coupling $g$.
We thus focus on the shape of the $E1$ dipole function by normalizing our result \eqref{dipole-strength-corrected} with Eq.~\eqref{eq:integrated-E1} and $B$ as
\begin{equation}\label{dipole-strength-normalized}
\frac{1}{N} \frac{\diff B(E1)}{\diff\omega} = 
 3 \theta(\omega-B)
 \frac{(\omega/B-1)^2}{(\omega/B)^4}
 \frac{f_{E1} \left( \frac{\beta}{\sqrt{\omega/B-1}} \right)
 + \frac{r_0}{a}f^{(1)}_{E1}\left(\frac{\beta}{\sqrt{\omega/B-1}}\right)}{ f_c(\beta) + \tilde r_0 f_c^{(1)}(\beta)},
\end{equation}
Figure~\ref{fig:B01} compares the normalized $E1$ dipole functions \eqref{dipole-strength-normalized} with and without the effective-range correction. 
As evident from Fig.~\ref{fig:B01}, incorporating the effective-range correction does not significantly alter the shape of the $E1$ dipole function.

For a more quantitative evaluation of the effective-range correction, we compute the relative correction to the $E1$ dipole strength, following from Eq.~\eqref{dipole-strength-corrected}, as
\begin{align}
  \label{eq:E1-relative-correction}
  \frac{\delta (\diff B(E1)/\diff\omega)}{(\diff B(E1)/\diff\omega)_0}
  =\frac{r_0}{a}
  \frac{f_{E1}^{(1)}\Big(\frac{\beta}{\sqrt{\omega/B-1}}\Big)}{f_{E1}\Big(\frac{\beta}{\sqrt{\omega/B-1}}\Big)} 
  \,.
\end{align}
The asymptotic behavior of this correction is given by
\begin{equation}
 \frac{\delta (\diff B(E1)/\diff\omega)}{(\diff B(E1)/\diff\omega)_0}= \begin{cases} \displaystyle{\frac38 r_0 a(\omega-B)}, & 0 < \omega-B \ll 1/a^2,\vspace{6pt} \\ \displaystyle{ \frac{r_0}{a}}\,,  & \omega-B \gg 1/a^2. \end{cases}
\end{equation} 
As shown in Fig.~\ref{fig:E1_relative_correction}, the magnitude of the relative correction is numerically small, remaining below $r_0/|a| \approx 0.15$ at all values considered.

\subsection{Electric polarizability}
\label{sec:electricalpolarizability}

As another indicator of the electromagnetic response, we compute the electric polarizability $\alpha_D$---a measure of how easily the halo nucleus can be deformed by external electric fields.

We first define the electric polarizability by considering the interaction energy between the dipole operator $\bm{\mathcal{M}}$ and an applied weak electric field $\bm{\mathcal{E}}$, given by
\begin{equation}
 H_{\mathrm{dipole}} 
 = - \sqrt{\frac{4\pi}{3}} \bm{\mathcal{M}} \cdot \bm{\mathcal{E}}
 = - Ze (\r_{\mathrm{core}} - \bm{R}_{\mathrm{cm}}) \cdot \bm{\mathcal{E}}.
\end{equation}
Let $|0'\>$ denote the perturbed halo ground state under the applied electric field.
The electric polarizability $\alpha_D$ is then defined as the linear coefficient of the induced dipole moment:
\begin{equation}
 \sqrt{\frac{4\pi}{3}} \< 0'| \bm{\mathcal{M}} |0'\> = \alpha_D \bm{\mathcal{E}}.    
\end{equation}
Using the standard formula for the perturbed ground state,
\begin{equation}
  |0'\> = |0\> + \sum_{n\neq0} c_n |n\>
  \quad \mathrm{with} \quad
  c_n 
  =  \sqrt{\frac{4\pi}{3}} \frac{\<n|\bm{\mathcal{M}}|0\>}{E_n - E_0} \cdot \bm{\mathcal{E}},
\end{equation}
we obtain an expression for the electric polarizability:
\begin{equation}
 \alpha_D 
 = {\frac13}\frac{4\pi}{3}
 \sum_{n\neq 0} \frac{2|\<n|\bm{\mathcal{M}}|0\>|^2}{E_n-E_0}.
\end{equation}
Furthermore, recalling Eq.~\eqref{eq:def-E1}, we derive another sum rule that relates the electric polarizability to the $E1$ dipole strength function:
\begin{equation}\label{eq:electric-pol-E1}
 \alpha_D 
 = \frac{2}{{3}}
 \frac{4\pi}{3} \int\limits_B^\infty \frac{\!\diff\omega}{\omega}\,\frac{\diff B(E1)}{\diff\omega} ,
\end{equation}

Substituting Eq.~\eqref{dipole-strength-corrected} and performing the above integral, we obtain the electric polarizability with the first-order effective-range correction as (see Appendix \ref{ap:electricalpolarizability} for details)
\begin{equation}\label{eq:electricalpolarizability}
  \alpha_\text{D} = (Ze)^2\frac{2g^2}{3\pi} \frac{A^{1/2}}{(A+2)^{5/2}}\frac1{B^2}
  \left[
   f_\text{pol}(\beta) - \tilde r_0\beta f_\text{pol}^{(1)}(\beta)
  \right] ,
\end{equation}
where we have introduced two dimensionless functions: 
\begin{align}
\label{eq:fpol0}
  f_{\mathrm{pol}} (\beta) 
  &=  \begin{cases}
  \displaystyle{\frac{\beta^2+2}{2(1-\beta^2)^2}
  -\frac{3\beta \arccos\beta}{2(1-\beta^2)^{5/2}}}, & \beta < 1,\vspace{6pt}\\
    \displaystyle{\frac15}\,, & \beta=1, \vspace{6pt} \\
  \displaystyle{\frac{\beta^2+2}{2(\beta^2-1)^2}
  -\frac{3\beta \arccosh\beta}{2(\beta^2-1)^{5/2}}}, & \beta > 1,
  \end{cases}
 \\
\label{eq:fpol1}
  f_{\mathrm{pol}}^{(1)}(\beta) 
  &=   
  \begin{cases}
  \displaystyle{\frac{11\beta^2+4}{4(1-\beta^2)^3}
  -\frac{3\beta(2\beta^2+3) \arccos\beta}{4(1-\beta^2)^{7/2}}}, & \beta < 1,\vspace{6pt}\\
    \displaystyle{\frac3{35}}\,, & \beta=1,\vspace{6pt} \\
   \displaystyle{-\frac{11\beta^2+4}{4(\beta^2-1)^3}
  +\frac{3\beta(2\beta^2+3) \arccosh\beta}{4(\beta^2-1)^{7/2}}}, & \beta > 1.
  \end{cases}
\end{align}
Note that the functions $f_{\textrm{pol}}(\beta)$ and $f_{\textrm{pol}}^{(1)}(\beta)$ are continuous across $\beta=1$. 
As in the previous sections, we provide alternative expressions for these functions in the regions $\beta<1$ and $\beta>1$, explicitly demonstrating that they remain real when their arguments are real. The asymptotic behavior of these functions at large and small $\beta$ is given by
\begin{equation}\label{eq:fpol-asymptotics}
    f_\text{pol}(\beta) \to \begin{cases}
    1 + O(\beta), 
    & \beta\ll 1, \vspace{6pt}\\
    \displaystyle{\frac{1}{2\beta^2} + O\left(\frac{\ln\beta}{\beta^4}\right)}, & \beta\gg 1,
    \end{cases}  \qquad f_\text{pol}^{(1)}(\beta) \to \begin{cases}
    1 + O(\beta), & \beta\ll 1,\vspace{6pt} \\
    \displaystyle{\frac{6\ln(2\beta)-11}{4\beta^4} + O\left(\frac{\ln\beta}{\beta^6}\right)}, & \beta\gg 1.
    \end{cases}
\end{equation}
The effective-range correction is suppressed by at least a factor of $\tilde r_0\beta = r_0/|a| \approx 0.15$.

\begin{figure}[t]
    \begin{minipage}{0.48\textwidth}
        \centering
        \includegraphics[width=\textwidth]{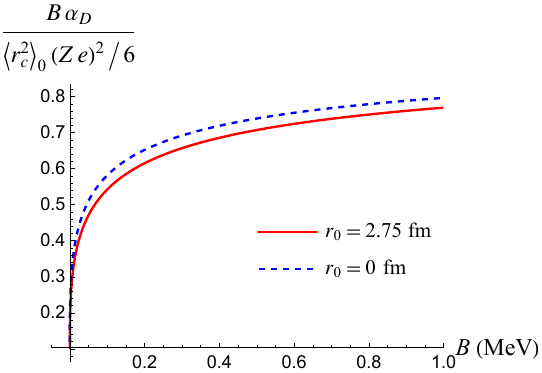}
    \caption{Ratio of electric polarizability and charge radius, Eq.\ \eqref{eq:alpha_rc}, with (solid) and without (dotted) the effective-range corrections as a function of $B$.}
    \label{fig:alpha_rc}
    \end{minipage}
\end{figure}

As the rms radii and the $E1$ dipole strength function, the electric polarizability is proportional to the running coupling $g^2$ and its absolute magnitude cannot be predicted in our EFT.  
On the other hand, one can construct a ratio involving the electric polarizability $\alpha_D$, charge radius, $\<r_c^2\>$ and binding energy $B$,
\begin{equation}
  \frac{B \alpha_\text{D}}{\<r_c^2\>} = \frac16 (Ze)^2
  \frac{f_\text{pol}(\beta) - \tilde r_0\beta f_\text{pol}^{(1)}(\beta)}{f_c(\beta)  + \tilde{r}_0 f_c^{(1)}(\beta)},
\label{eq:alpha_rc}
\end{equation}
which we can can predict.
This ratio has a particularly simple form in the unitarity regime of neutron-neutron interaction ($a=\infty$, $r_0=0$, or $\beta \to 0,\tilde{r}_0=0$):
\begin{equation}
  \frac{B \alpha_\text{D}}{\<r_c^2\>}\bigg|_\text{unitarity} = \frac16 (Ze)^2 .
  \label{eq:alpha_rc-unitarity}
\end{equation}
At the physical scattering length the value of the ratio is less than the unitarity value.  
The correction due to the effective range is rather small, as seen in Fig.\ \ref{fig:alpha_rc}.

\section{Concluding remarks}
\label{sec:conclusion}

In this paper, we evaluated the corrections from the neutron-neutron effective range to various physical quantities of two-neutron halo (Borromean) nuclei.
We computed the effective-range correction to the mean-square radii, demonstrating its contributions to the ratio of the matter and charge radii, as well as the magnitude of the relative corrections to the zero-range result.
Additionally, we evaluated the effective-range corrections to the electromagnetic response, including the $E1$ dipole strength function and the electric polarizability.
We found that for many physical quantities, corrections are suppressed by at least a factor of $r_0/|a| \approx 0.15$.
The only quantity where the correction is relatively large is the inter-neutron distance $\<\r^2\>$ (see Fig.~\ref{fig:r-relative-correction}).
For this observable, the correction is already of order 1 when the binding energy of the halo $B$ is around $1$ MeV.  
This suggests that a resummation of the effective-range corrections may be necessary for this particular quantity.

There are several potential directions worth pursuing based on the present work.

One direction is to compute corrections arising from other sources.
As mentioned in Sec.~\ref{sec:Lagrangian}, two leading corrections---aside from the neutron-neutron effective range considered in this paper---are accessible within the framework of effective field theory.
The first source of correction involves incorporating the leading irrelevant operator responsible for the core-neutron scattering length, which contributes to Feynman diagrams with three-loop corrections.
The second source arises in cases where a core-neutron $p$-wave resonance is present, whose effects can be treated perturbatively following the method of Ref.~\cite{Bertulani:2002sz}.

Another interesting direction is to explore the correspondence with other formulations, such as the first-quantized approach.
For instance, one may expect that the effective-range correction to the ratio $\<r_m^2\>/\<r_c^2\>$ can be obtained using the alternative method of Ref.~\cite{Naidon:2023fio}, where the three-body wave function remains dominated by a single Faddeev component.
On the other hand, an EFT-based computation of the wave function should also be possible.
A direct comparison between our EFT approach and first-quantized methods (such as those based on the Faddeev equation) would be a valuable subject for further investigation.
We defer these interesting directions to future work.

\begin{acknowledgments}

The authors thank Pascal Naidon for stimulating discussion. The authors thank the Tohoku Forum for Creativity at Tohoku University. Discussions during the junior research program ``Universality of Strongly Correlated Few–body and Many–body Quantum Systems" were useful to complete this work.  This work is supported, in part, by U.S.\ DOE Grant No.\ DE-
FG02-01ER41195.  
DBC acknowledge support from the Simons Collaboration on Global Categorical Symmetries, the US Department of Energy Grant 5-29073, and the Sloan Foundation.
MH is supported by the Japan Society for the Promotion of Science (JSPS) KAKENHI Grant No.\ JP22K20369 and No.\ JP23H01174.

\end{acknowledgments}

\appendix

\section{A simple model of a two-neutron halo nucleus without core-halo resonance}
\label{app:variational}

In this Appendix, using a simple quantum mechanical model, we demonstrate that a bound state of a core and two neutrons can exist even when only the neutron-neutron interaction is resonant, while the core-neutron interaction remains far from resonance.

Our model consists of a system of two particles (``neutrons") and a third, infinitely massive particle (``core").
In contrast to the EFT approach in the main text, the dynamics of the heavy core particle is assumed to be suppressed, serving instead as a fixed source of the neutron-core interaction at its position, $\r_{\mathrm{core}} =0$.
Modeling this non-resonant neutron-core interaction as an attractive Gaussian potential, we consider the Hamiltonian given by
\begin{equation}
  H = -\frac12 \nabla_{\r_1}^2 - \frac12 \nabla_{\r_2}^2
  - V_0 \left( e^{-r_1^2/2} + e^{-r_2^2/2} \right) ,
\end{equation}
where $\r_1$ and $\r_2$ denote the positions of the two neutrons.

On the other hand, the two neutrons interact with each other through a resonant point-like interaction.
This interaction is implemented by imposing the Bethe-Peierls boundary condition on the wave function~\cite{Bethe-Peierls1935}.
To express this condition, we introduce the Jacobi coordinates [corresponding to Eqs.~\eqref{eq:inter-neutron-dis}-\eqref{eq:Jacobi-rho} with $A \to \infty$]:
\begin{equation}
  \bm\rho = \r_1+\r_2\,, \qquad \r = \r_1 - \r_2 .
\end{equation}
and require that the $s$-wave component of the wave function be a singular function of $\r$ as the inter-neutron distance approaches zero ($\r\to0$):
\begin{equation}\label{eq:B-P}
 \psi(\bm\rho, \r) \sim \frac{C(\bm\rho)}r + O(r) + (\text{higher partial waves in $\r$}).
\end{equation}

We now show that this model could support a three-body bound state despite the non-resonant core-neutron interaction.
To this end, we first note that the two-particle subsystem of the core and one neutron forms a bound state when the attraction is sufficiently strong.
Indeed, solving the Schr\"odinger equation numerically shows that such a bound state appears when $V_0$ exceeds the critical value $V_0^{cn}\approx0.671$. 
We then define $V_0^\text{3bdy}$ as the minimum value required for the formation of a three-body bound state.
Below, using the variational method, we demonstrate that $V_0^\text{3bdy} < V_0^{cn}\approx0.671$.
In other words, we establish the existence of a window,  $V_0^\text{3bdy} < V_0 < V_0^{cn}\approx0.671$, in which the attractive potential is too weak to form a two-body bound state but strong enough to support a three-body bound state.

To estimate $V_0^\text{3bdy}$, we recall that the energy expectation value for any variational wave function is always an upper bound on the true ground state energy.
In particular, a three-body bound state appears when the variational energy becomes negative for $V_0 < V_0^{cn} (\approx0.671)$.
We determine the minimum $V_0$ at which the variational energy reaches zero using two different variational wave functions, thereby confirming that $V_0^\text{3bdy} < V_0^{cn} (\approx0.671)$.

As the first simple variational wave function, we take
\begin{equation}
  \psi(\bm\rho,\r) = 
  \frac{\alpha}{2\pi^{3/2}} \frac{\rme^{-\alpha(r_1^2+r_2^2)/4}}r =
  \frac\alpha{2\pi^{3/2}} e^{-\alpha 
  \rho^2/8}\,
  \frac{e^{-\alpha r^2/8}}r \,,
\end{equation}
which corresponds to the ground-state wave function of two resonantly interacting particles in a harmonic potential with an appropriately chosen oscillator frequency. This wave function explicitly satisfies the Bethe-Peierls boundary condition.
Using this variational wave function, the energy expectation value is given by
\begin{equation}
  E(\alpha) = \frac\alpha2 - \frac{4V_0\alpha^{3/2}}{(1+2\alpha)
    \sqrt{1+\alpha}}\,.
\end{equation}
Minimizing $E(\alpha)$ for a given $V_0$, we obtain an upper bound on the ground-state energy of the three-body system. This upper bound reaches zero at $V_0=\frac18 \phi (2\phi+1)^{1/2}\approx 0.416$, where $\phi$ is the golden ratio [At this value of $V_0$, the minimum of $E(\alpha)$ occurs at $\alpha=(\phi-1)/2$)].  
Our variational calculation thus establishes that a bound state must exist for $V_0>0.416$, providing an upper bound on the critical potential depth: $V_0^\text{3bdy}<0.416$.  
Notably, this upper bound is less than $2/3$ of $V_0^{cn}$, the critical value at which the two-body bound state appears.

With an alternative variational approach, we can further refine our estimate.
First, we express the Hamiltonian in terms of the Jacobi coordinates:
\begin{equation}
  H = -  \frac{\d^2}{\d\r^2} - \frac{\d^2}{\d\bm{\rho}^2}
  - V_0 \exp\left( -\frac{(\r+\bm{\rho})^2}8 \right)
  - V_0 \exp\left( -\frac{(\r-\bm{\rho})^2}8 \right).
\end{equation}
Next, changing to the coordinates $R$, $\alpha$, $\hat\r$, and $\hat{\bm{\rho}}$ where $\hat r$ and $\hat{\bm{\rho}}$ denote the directions of $\r$ and $\bm{\rho}$, respectively, and
\begin{align}
  r = R\sin\alpha , \quad 
  \rho = R \cos\alpha ,
\end{align}
we rewrite the Hamiltonian as
\begin{equation}
 \begin{split}
  H =& - \frac{\d^2}{\d R^2} - \frac5R\frac\d{\d R}
  - \frac1{R^2}\frac{\d^2}{\d\alpha^2}
  - \frac2{R^2} \left( \frac1{\tan\alpha} -\tan\alpha \right)\frac\d{\d\alpha}
  \\
  &- \frac1{R^2\sin^2\alpha}\frac{\d^2}{\d\hat\r^2}
  - \frac1{R^2\cos^2\alpha}\frac{\d^2}{\d\hat{\bm\rho}^2}
  - 2V_0 e^{-R^2/8} \cosh\left( \frac{R^2}8\sin2\alpha
  \,\hat\r\cdot\hat{\bm\rho} \right).
 \end{split}
\end{equation}
To construct a variational wave function, we restrict our consideration to isotropic wave functions in $\hat\r$ and $\hat{\bm\rho}$.  
Within this class, the last term in the Hamiltonian can be replaced by its angular average, leading to the effective form
\begin{equation}
  H = - \frac{\d^2}{\d R^2} - \frac5R\frac\d{\d R}
  - \frac1{R^2}\frac{\d^2}{\d\alpha^2}
  - \frac2{R^2} \left( \frac1{\tan\alpha} -\tan\alpha \right)\frac\d{\d\alpha}
  - 2V_0 e^{-R^2/8} \frac{\sinh\left( \frac{R^2}8\sin2\alpha\right)}
    {\frac{R^2}8\sin2\alpha}\,.
\end{equation}
We further assume a separable ansatz for the wave function:
\begin{equation}
  \psi(R,\alpha) = \frac{\psi(R)}{\sin\alpha} \,.
\end{equation}
which remains consistent with the Bethe-Peierls boundary condition~(\ref{eq:B-P}). 
Applying this ansatz simplifies the Hamiltonian to
\begin{equation}
  H = - \frac{\d^2}{\d R^2} - \frac5R\frac\d{\d R} - \frac3{R^2}
  - 2V_0 e^{-R^2/8}
  \left\langle \frac{\sinh\left( \frac{R^2}8\sin2\alpha\right)}
    {\frac{R^2}8\sin2\alpha} \right\rangle_\alpha ,
\end{equation}
where $\<\>_\alpha$ denotes the average over $\alpha$:
\begin{equation}
  \< f(\alpha)\>_\alpha \equiv \frac4\pi \!
  \int_0^{\pi/2} \!d\alpha\, \cos^2\alpha\, f(\alpha).
\end{equation}
To compute the variational energy, we expand the wave function in a Taylor series in $\alpha$, average each term, and then resum the series. This procedure yields the effective Hamiltonian
\begin{equation}
  H = - \frac{\d^2}{\d R^2} - \frac5R\frac\d{\d R} - \frac3{R^2}
  - 2V_0 e^{-R^2/8}
  {}_p F_q\left( \frac12; 1, \frac32; \frac{R^4}{256} \right).
\end{equation}
Solving the Schrödinger equation numerically, we find that this Hamiltonian develops a bound state at $V_0\approx 0.3285$, providing an improved estimate for $V_0^\text{3bdy}$: $V_0^\text{3bdy}<0.3285$. 
Thus, for a Gaussian-shaped potential, the critical depth required to bind two resonantly interacting particles is less than half the critical depth needed to bind a single particle.

Another way to quantify how far the core-neutron interaction is from resonance is to compute the scattering length and effective range for low-energy neutron scattering on a Gaussian potential with depth $V_0=0.3285$.
At this depth, we find the scattering length $a\approx-1.570$ and effective range $r_0\approx3.324$.  
Clearly, the hierarchy $|a|\gg r_0$ which one would expect if the neutron-core interaction were near resonance, is not satisfied.

\section{Details of calculations}\label{app:details}

In this Appendix we provide the detailed computations omitted in the main text.
We present the calculations of the charge radius (Sec.~\ref{ap:Charge-radius}), the dineutron and matter radii (Sec.~\ref{ap:Neutron-Matter-radius}), the $E1$ dipole strength (Sec.~\ref{ap:dipole-strength-corrected}), and the electric polarizability (Sec.~\ref{ap:electricalpolarizability}) in sequence.

\subsection{Charge radius}
\label{ap:Charge-radius}

We here present the detailed computation of the charge radius corrections presented in Section \ref{sec:Charge-radius}.
The integral expression for the charge radius is given by Eq.\eqref{eq:chargeradius}.
Changing the integration variable from $|\q|$ to $y = B_\q / B$ with $B_q$ defined in Eq.~\eqref{eq:Bq} and taking into account that $m_\phi=A$, $m_h=A+2$, and $\mu=2A/(A+2)$, we obtain
\begin{equation}
    \langle r_c^2\rangle=\frac{4g^2}{\pi B}\frac{A^{1/2}}{(A+2)^{5/2}}f_c(\beta,\tilde{r}_0),
\end{equation}
with the dimensionless function of the parameters defined in Eq.~\eqref{betarho}:
\begin{equation}\label{fcbetar0}
 \begin{split}
  f_c(\beta,\tilde r_0) 
  &= \int\limits_1^\infty\! \diff y\, \sqrt{y-1} 
  \left[  3 f''(y;-1/\beta,\tilde r_0)
  + \frac23 (y-1) f'''(y;-1/\beta,\tilde r_0) \right]
  = - \int\limits_1^\infty\! \diff y\,  \frac{f'(y;-1/\beta,\tilde r_0)}{\sqrt{y-1}}\, ,
 \end{split}
\end{equation}
where we performed integration by parts to obtain the rightmost expression.

Expanding $f'(y;-1/\beta,\tilde r_0)$  in terms of the small parameter characterizing the effective range, $\tilde r_0 = r_0 \sqrt{B}$, and performing the integrals for each term, we derive 
\begin{align}
  f_c(\beta,\tilde{r}_0)=f_c(\beta)+\tilde{r}_0f_c^{(1)}(\beta)+O(\tilde{r}_0^2),
\end{align}
with the analytic expressions for $f_c(\beta)$ and $f_c^{(1)}(\beta)$:
\begin{align}
  f_c(\beta) = -\!\int\limits_1^\infty\! \diff y\, \frac{\partial_y f(y;-1/\beta)}{\sqrt{y-1}} 
  &= \begin{cases}
    \displaystyle{\frac1{1-\beta^2}} - \displaystyle{\frac{\beta\arccos\beta}{(1-\beta^2)^{3/2}}} \,, &     \beta<1,\vspace{6pt} \\
    \displaystyle{\frac13}\,, &\beta =1, \vspace{3pt} \\
    \displaystyle{-\frac1{\beta^2-1}} + \displaystyle{\frac{\beta\arccosh\beta}{(\beta^2-1)^{3/2}}}\,, &\beta>1,
    \end{cases}
 \\
  f_c^{(1)}(\beta) 
  = - \frac 12 \!\int\limits_1^\infty\! \diff y\,
  \frac{ \d_y [y f^2(y;-1/\beta)]}{\sqrt{y-1}} 
  &= \begin{cases}
  \displaystyle{-\frac\beta2 \left[ \frac{2+\beta^2}{(1-\beta^2)^2} - \frac{3\beta\arccos\beta}{(1-\beta^2)^{5/2}} \right]},  & \beta <1,\vspace{6pt}\\
  -\displaystyle{\frac15}\,,& \beta = 1,\vspace{6pt} \\
  \displaystyle{-\frac\beta2 \left[ \frac{2+\beta^2}{(\beta^2-1)^2} - \frac{3\beta\arccosh\beta}{(\beta^2-1)^{5/2}} \right]},  & \beta >1,
  \end{cases}
\end{align}
which are Eqs.~\eqref{eq:fc0} and \eqref{eq:fc1} in the main text.

\subsection{Dineutron radius}
\label{ap:Neutron-Matter-radius}

We here present the detailed computation of the dineutron radius presented in Section \ref{sec:Neutron-Matter-radius}.
The integral expression for the dineutron radius is given by Eq.~\eqref{eq:rn}. 
Performing the angular integral and changing the radial integration variable to $y= B_\q/B$, we obtain 
\begin{align}
    \begin{split}
        \< r_n^2\> = - \frac{g^2}{4\pi} \frac{(2\mu)^{3/2}}B \!
  \int\limits_1^\infty \! \diff y\,
  \sqrt{y-1}\,
  \bigg\{
  & \bigg[ \bigg( -\frac5{8} + \frac3{4m_h}\bigg)\frac1{y^{3/2}}
  + \frac{2\mu(y-1)}{32 y^{5/2}}
  \bigg]f^2 \bigg. \\
  \bigg. & + \left( \frac1{\sqrt{y}} - \tilde r_0 \right)
  \bigg[ \frac{3m_\phi}{2m_h} f\! f'
  + \frac{2\mu(y-1)}{4} (f\! f''-f'^2) \bigg]
  \bigg\} ,
    \end{split}
\end{align}
where, in this section, we use the simplified notation $f\equiv f(y;-1/\beta,\tilde r_0)$.
Using $m_\phi=A$, $m_h=A+2$, and $\mu=2A/(A+2)$,  we then express the dineutron radius as
\begin{equation}
  \< r_n^2 \> = \frac{g^2}{\pi B}
  \left( \frac{A}{A+2} \right)^{3/2}
  \left[ f_n(\beta,\tilde r_0) + \frac A{A+2} f_c(\beta,\tilde r_0) \right],
\end{equation}
where we introduce two dimensionless functions
\begin{align}
  f_n(\beta, \tilde r_0) 
  &= \int\limits_1^\infty \!\diff y\,
  \frac{\sqrt{y-1}}{2y^{3/2}} f^2,
  \label{eq:fn2}
  \\
  f_c(\beta,\tilde r_0)  
  &= \int\limits_1^\infty \! \diff y\, \sqrt{y-1}
  \,\Bigg\{ \!
  \left( \frac3{4y^{3/2}} - \frac{y-1}{4y^{5/2}}
  \right) f^2
  - \left( \frac1{\sqrt{y}} - \tilde r_0 \right)
  \left[ 3 f\! f'
  + 2(y-1) (f\! f''-f'^2) \right] \!
  \Bigg\}.
  \label{eq:fc2}
\end{align}
One can verify that the following nontrivial identity holds:
\begin{equation}
   \left( \frac3{4y^{3/2}} - \frac{y-1}{4y^{5/2}}
  \right) f^2 
   - \left( \frac1{\sqrt{y}} - \tilde r_0 \right)
  \left[ 3 f\!f'
  + 2(y-1) (f\!f''-f'^2) \right]
  = 3 f'' + \frac23 (y-1)f''',
\end{equation}
which allows us to show that the integral on the right-hand side of Eq.~\eqref{eq:fc2} yields the same function as the one defined in Eq.~(\ref{fcbetar0}).

We then expand Eq. \eqref{eq:fn2} as a power series in $\tilde{r}_0$, resulting in
\begin{equation}
  f_n(\beta,\tilde r_0) = f_n(\beta) + \tilde{r}_0 f_{n}^{(1)}(\beta)+O(\tilde{r}_0^2),
\end{equation}
with the following analytic expression:
\begin{align}
  f_n (\beta) = 
  \int\limits_1^\infty \diff y\,
  \frac{\sqrt{y-1}}{2y^{3/2}} \left( \frac1{\sqrt y+\beta}\right)^2 
  &=\begin{cases}
     \displaystyle{\frac1{\beta^3}} \biggl[ \pi-2\beta +
    (\beta^2-2)\displaystyle{
    \frac{\arccos\beta}{\sqrt{1-\beta^2}} }\biggr], & \beta <1,\vspace{4pt}\\
    \pi-3, & \beta =1, \vspace{4pt}\\
    \displaystyle{\frac1{\beta^3}} \biggl[  \pi-2\beta +
    (\beta^2-2)\displaystyle{
    \frac{\arccosh\beta}{\sqrt{\beta^2-1}} }\biggr], & \beta>1,
    \end{cases}
 \\
  f_n^{(1)}(\beta) 
  = \int\limits_1^\infty \diff y\,
 \frac{\sqrt{y-1}}{2y^{1/2}} \left( \frac1{\sqrt y+\beta}\right)^3 
  &= \begin{cases}
     \displaystyle{\frac12 \left[ - \frac\beta{1-\beta^2}
     + \frac{\arccos\beta}{(1-\beta^2)^{3/2}}
     \right]}, & \beta < 1, \vspace{6pt}\\
     \displaystyle{\frac13}\,, & \beta=1,\vspace{6pt} \\
     \displaystyle{\frac12 \left[ \frac\beta{\beta^2-1}
     - \frac{\arccosh\beta}{(\beta^2-1)^{3/2}}
     \right]}, & \beta > 1,
  \end{cases}
\end{align}
which corresponds to Eqs.~\eqref{eq:fn0} and \eqref{eq:fn1} in the main text.

\subsection{\texorpdfstring{$E1$}{E1} dipole strength function}
\label{ap:dipole-strength-corrected}

We here we present the detailed computation of the $E1$ dipole strength function presented in Section \ref{sec:E1-dipole}.
The $E1$ dipole strength function is given by Eq.~\eqref{eq:E1-fcn}.
We first perform the angular part of the integral and use the Heaviside theta function to rewrite it as
\begin{equation}
  \Im G_{JJ}(\omega) = - (Ze)^2 \frac{2g^2}{\pi m_\phi^2\omega^2}\!
  \int\limits_0^{\sqrt{2\mu(\omega-B)}}\!\diff q\, q^4
  \frac{\sqrt{\omega-B-\frac{\q^2}{2\mu}}}
  {\omega-B-\frac{\q^2}{2\mu} + \left[\frac1a -\frac{r_0}2
  \left(\omega-B-\frac{\q^2}{2\mu}\right) \right]^2}.
\end{equation}
Next, we change the radial integration variable from $q=|\q|$ to $u$, defined by $q = \sqrt{2\mu} \sqrt{\omega-B} \sqrt u$, which results in
\begin{equation}
 \begin{split}
  \Im G_{JJ}(\omega) 
  &= - (Ze)^2 \frac{g^2}{\pi m_\phi^2\omega^2}
  (2\mu)^{5/2} (\omega-B)^{5/2} \!
  \int\limits_0^1\!\diff u\, u^{3/2}
  \frac{\sqrt{(\omega-B)(1-u)}}
  {(\omega-B)(1-u) + \left[\frac1{-a} +\frac{r_0}2
  (\omega-B)(1-u) \right]^2}
  \\
  &= - (Ze)^2 \frac{g^2(2\mu)^{5/2}}{\pi m_\phi^2}
   \frac{(\omega-B)^2}{\omega^2} \!
  \int\limits_0^1\!\diff u\, \left[
    \frac{u^{3/2}\sqrt{1-u}}{1-u + x^2}
  + \frac{r_0}{a} \frac{u^{3/2}(1-u)^{3/2}}{(1-u+x^2)^2}  
  \right]+O(r_0^2),
 \end{split}
\end{equation}
where, in the second line, we expanded in powers of $r_0/a$ and introduced $x=1/(-a\sqrt{\omega-B})$. 
Substituting this expression, we obtain
\begin{equation}
  \frac{\diff B(E1)}{\diff\omega} = \frac3{4\pi}(Ze)^2
  \frac{32g^2}{\pi^2} \frac{A^{1/2}}{(A+2)^{5/2}} \frac{(\omega-B)^2}{\omega^4}
   \int\limits_0^1\!\diff u\, \left[
    \frac{u^{3/2}\sqrt{1-u}}{1-u + x^2}
  + \frac{r_0}{a} \frac{u^{3/2}(1-u)^{3/2}}{(1-u+x^2)^2}  
  \right]+O(r_0^2),
\end{equation}
We can readily compute the two integrals appearing above as
\begin{align}
 \int\limits_0^1\!\diff u\, 
 \frac{u^{3/2}\sqrt{1-u}}{1-u + x^2} 
 &= \frac{3\pi}8 
 \left[ 1-\frac{8}{3}x(1+x^2)^{3/2}+4x^2\Big(1+\frac{2}{3}x^2\Big)
 \right]
 \equiv \frac{3\pi}8 f_{E1}(x)
 ,\\
  \int\limits_0^1\!\diff u\,  \frac{u^{3/2}(1-u)^{3/2}}{(1-u+x^2)^2}
  &= \frac{3\pi}{8} 
  \left[ 1 - 4x(1+2x^2)\sqrt{1+x^2} + 8 x^2 (1+x^2) \right]
  \equiv \frac{3\pi}8 f^{(1)}_{E1}(x),
\end{align}
where we defined $f_{E1}(x)$ and $f_{E1}^{(1)}(x)$ as given in Eqs.~\eqref{fE1} and \eqref{fE11} of the main text.
As a result, we obtain the final expression for the $E1$ dipole strength, including the first-order effective-range corrections, as
\begin{equation}
  \frac{\diff B(E1)}{\diff\omega} = \frac3{4\pi}(Ze)^2
  \frac{12g^2}{\pi} \frac{A^{1/2}}{(A+2)^{5/2}}\theta(\omega-B)
  \frac{(\omega-B)^2}{\omega^4}
  \left[ f_{E1} \left( \frac1{-a\sqrt{\omega-B}} \right)
  + \frac{r_0}{a}f^{(1)}_{E1}\left(\frac1{-a\sqrt{\omega-B}}\right)\right],
\end{equation}
which corresponds to Eq.~\eqref{dipole-strength-corrected} in the main text.

\subsection{Computing the electrical polarizability}
\label{ap:electricalpolarizability}

We here present the detailed computation of the electric polarizability presented in Section \ref{sec:electricalpolarizability}.
As shown in the main text, the electric polarizability is expressed as a weighted frequency sum of the $E1$ dipole strength, given by Eq.~\eqref{eq:electric-pol-E1}.
Substituting the expression of the $E1$ dipole strength from Eq.~(\ref{dipole-strength-corrected}) and rescaling the integration variable by $\omega=By$, we obtain
\begin{equation}
  \alpha_D = (Ze)^2\frac{8g^2}\pi \frac{A^{1/2}}{(A+2)^{5/2}}\frac1{B^2} \!
  \int\limits_1^\infty\! \diff y\, \frac{(y-1)^2}{y^5} \left[
  f_{E1}\left(\frac\beta{\sqrt{y-1}}\right)
  + \frac{r_0}{a} f_{E1}^{(1)}\left(\frac\beta{\sqrt{y-1}}\right)
  \right].
\end{equation}
The two integrals can be computed directly:
\begin{align}
  \int\limits_1^\infty\!\diff y\, \frac{(y-1)^2}{y^5}
  f_{E1} \left( \frac\beta{\sqrt{y-1}} \right) 
  &= \frac1{12} f_{\text{pol}}(\beta),\\
  \int\limits_1^\infty\!\diff y\, \frac{(y-1)^2}{y^5}
  f^{(1)}_{E1} \left( \frac\beta{\sqrt{y-1}} \right) &=
  \frac1{12} f^{(1)}_{\text{pol}}(\beta),
\end{align}
where we introduced
\begin{align}
  f_{\mathrm{pol}} (\beta) 
  &=  \begin{cases}
  \displaystyle{\frac{\beta^2+2}{2(1-\beta^2)^2}
  -\frac{3\beta \arccos\beta}{2(1-\beta^2)^{5/2}}}, & \beta < 1,\vspace{6pt}\\
    \displaystyle{\frac15}\,, & \beta=1, \vspace{6pt} \\
  \displaystyle{\frac{\beta^2+2}{2(\beta^2-1)^2}
  -\frac{3\beta \arccosh\beta}{2(\beta^2-1)^{5/2}}}, & \beta > 1,
  \end{cases}
 \\
  f_{\mathrm{pol}}^{(1)}(\beta) 
  &=   
  \begin{cases}
  \displaystyle{\frac{11\beta^2+4}{4(1-\beta^2)^3}
  -\frac{3\beta(2\beta^2+3) \arccos\beta}{4(1-\beta^2)^{7/2}}}, & \beta < 1,\vspace{6pt}\\
    \displaystyle{\frac3{35}}\,, & \beta=1,\vspace{6pt} \\
   \displaystyle{-\frac{11\beta^2+4}{4(\beta^2-1)^3}
  +\frac{3\beta(2\beta^2+3) \arccosh\beta}{4(\beta^2-1)^{7/2}}}, & \beta > 1.
  \end{cases}
\end{align}
Thus, the final result for the electric polarizability with first-order effective-range corrections is
\begin{equation}
  \alpha_D = (Ze)^2\frac{2g^2}{3\pi} \frac{A^{1/2}}{(A+2)^{5/2}}\frac1{B^2}
  \left[ f_\text{pol}(\beta) + \frac{r_0}{a}f_\text{pol}^{(1)}(\beta)\right],
\end{equation}
which corresponds to Eq.~\eqref{eq:electricalpolarizability} in the main text.

\bibliography{effective_range}

\end{document}